\documentclass{revtex4}
\usepackage{graphicx}
\usepackage{subfigure,psfrag}
\newcommand{\etal}{{\it et al.}~}
\newcommand{\bs}{\mathbf {s}}
\newcommand{\br}{\mathbf {r}}
\newcommand{\bv}{\mathbf {v}}

\begin{document}

\title{Tree method for quantum vortex dynamics}


\author{A.~W.~Baggaley}
\email{a.w.baggaley@ncl.ac.uk}
\author{C.~F.~Barenghi} 
\email{c.f.barenghi@ncl.ac.uk}
\affiliation{School of Mathematics and Statistics, University of
Newcastle, Newcastle upon Tyne, NE1 7RU, UK}




\begin{abstract}
We present a numerical method to compute the evolution of vortex
filaments in superfluid helium. The method is based on a tree
algorithm which considerably speeds up the calculation of Biot-Savart
integrals. We show that the computational cost scales as $N\log{(N)}$ 
rather than $N^2$, where
$N$ is the number of discretization points. We test the method and
its properties for a variety of vortex configurations, ranging from simple
vortex rings to a counterflow vortex tangle, and compare results against
the Local Induction Approximation and the exact Biot-Savart law.
\end{abstract}

\maketitle

\section{Introduction}

Quantum turbulence \cite{Donnelly,Barenghi-Sergeev,Halperin}
consists of a disordered tangle of
reconnecting quantized vortex filaments, each carrying one quantum
of circulation $\kappa$.  This form of turbulence is easily created 
by agitating superfluid liquid helium  ($^4$He) 
with propellers \cite{Tabeling,Roche2007}, forks\cite{Skrbek}, or
grids \cite{Smith1993}, the same techniques which are used to create
turbulence in ordinary fluids. There are also methods which are unique to
superfluid liquid helium, for example heat flows \cite{Vinen1957,Paoletti2008}
and streams of ions \cite{Walmsley2008}. Quantum turbulence is also studied 
in superfluid $^3$He-B \cite{Eltsov2010,Lancaster2011} and, more recently,
in atomic Bose-Einstein condensates\cite{Henn2009}. In the study of
quantum turbulence, particular attention
is devoted to the similarities with
ordinary turbulence \cite{Vinen-Niemela,Lvov-Nazarenko}.
Recent progress has been boosted by the development of new flow visualization 
techniques, such  as
tracer particles\cite{Maryland-tracers,VanSciver}, Andreev scattering
\cite{Lancaster-Andreev}
and laser-induced fluorescence\cite{Yale}.

This article is concerned with numerical simulations of quantum
turbulence. 
Starting from the pioneering work of Schwarz \cite{Schwarz},
numerical simulations have always played an important role 
in the field 
\cite{Samuels,Aarts,Bauer,Tsubota2003,Risto,Konda,Kivo,Morris,Kivo2,cascade,tree}.
The recent experimental progress, rather than reducing the need
of numerical simulations, has highlighted their importance 
in interpreting visualization methods 
\cite{Kivotides-PIV,Finne}. 

In superfluid $^4$He, the vortex core radius 
($a_0 \approx 10^{-8}~\rm cm$) is many orders of magnitude smaller than the
average separation between vortex lines (typically from
$10^{-2}$ to $10^{-4} \rm cm$) or any other relevant length scale in the flow. 
This feature was early recognised by Schwarz\cite{Schwarz},
who proposed to model vortex lines as spaces curves $\bs=\bs(\xi,t)$ of
infinitesimal thickness, where $t$ is the time and $\xi$ is arc length.
This vortex filament approach
is also valid in $^3$He-B, although to a lesser extent
because the core size is larger ($a_0 \approx 10^{-6}~\rm cm$).
The space curves are numerically discretized by a large, variable number of
points $\bs_i$ ($i=1,\cdots N)$, which hereafter we refer to as vortex points.

At temperatures low enough that the
normal fluid and mutual friction is negligible, the governing equation of
motion of the superfluid
vortex lines is the Biot-Savart (BS) law \cite{Saffman}

\begin{equation}
\frac{d\bs}{dt}=-\frac{\kappa}{4 \pi} \oint_{\cal L} \frac{(\bs-\br) }
{\vert \bs - \br \vert^3}
\times {\bf d}\br,
\label{eq:BS}
\end{equation}

\noindent
where the line integral extends over the entire vortex configuration
$\cal L$.  Eq.~(\ref{eq:BS}) formulates the classical Euler dynamics of
vortex filaments in
incompressible inviscid flows. The assumption of
incompressibility is justified, as all relevant speeds, 
of the order of few cm/s, are much smaller than
the speed of sound, which is approximately $240~\rm m/s$ 
(but see the discussion of reconnections and dissipation in
Section \ref{sec:algo}).
Unlike vortex lines in classical inviscid Euler flows, however,
superfluid vortex lines can reconnect with each other
when they come sufficiently close \cite{Koplik,Bewley,Tebbs,Kerr,Bajer}. This
effect is captured by a more microscopic model,
the Gross-Pitaevskii Equation (GPE) for a Bose-Einstein
condensate\cite{infinite}. In the context of the vortex filament
approach, reconnections are modelled by supplementing Eq.~(\ref{eq:BS}) 
with an algorithmical reconnection procedure, as originally
proposed by Schwarz \cite{Schwarz}.

Two difficulties arise in seeking a numerical solution of Eq.~(\ref{eq:BS}).
Firstly Eq.~(\ref{eq:BS}) is singular at $\bs=\br$.
Following Schwarz \cite{Schwarz},
to circumvent this problem we split the velocity of the filament 
of the $i^\textrm{th}$ vortex point $\bs_i$ into a local contribution 
(near the singularity) and non-local contributions, writing

\begin{equation}
\frac{d\bs_i}{dt}=
\frac{\kappa}{4\pi} \ln \left(\frac{\sqrt{\ell_i \ell_{i+1}}}{a_1}\right)\bs_i'   \times \bs_i'' 
-\frac{\kappa}{4 \pi} \oint_{\cal L'} \frac{(\bs_i-\br) }
{\vert \bs_i - \br \vert^3}
\times {\bf d}\br.
\label{eq:BS_desing}
\end{equation}

\noindent
Here a prime denotes a derivative with respect to arc length 
(hence $\bs'=d\bs/d\xi$ and $\bs''=d^2\bs/d\xi^2$),
$\ell_i$ and $\ell_{i+1}$ are the arclengths of the curves
between points $\bs_{i-1}$ and $\bs_i$ and between $\bs_i$ and $\bs_{i+1}$,
$\cal L'$ is the original vortex configuration $\cal L$ 
 without the section between $\bs_{i-1}$ and $\bs_{i+1}$,
and $a_1$ is a cutoff parameter.
For simplicity, hereafter we assume that the cutoff parameter is 
equal to the vortex core radius, $a_1\approx a_0$.

To numerically calculate the nonlocal term in Eq.~(\ref{eq:BS_desing}), we must perform further manipulations. 
We follow \cite{Schwarz} and split the integral into discrete
parts, which account for the contribution of the vortex between points along $\cal L'$.
Assuming a straight line between vortex points, the contribution to the nonlocal term due to the vortex segment between points $\bs_j$ and $\bs_{j+1}$ is given by,
\begin{equation}\label{eq:BS_schwarz_nl}
\Delta v_{nl}^{j}(\bs_i)=\frac{\kappa}{2\pi(4AC-B^2)}\left[ \frac{2C+B}{\sqrt{A+B+C}}-\frac{B}{\sqrt{A}}\right]\mathbf{p} \times \mathbf{q}.
\end{equation}
Where $\mathbf{p}=\bs_j-\bs_i$, $\mathbf{q}=\bs_{j+1}-\bs_j$, $A=|\mathbf{p}|^2$, $B=2\mathbf{p}\cdot \mathbf{q}$ and $C=|\mathbf{q}|^2$.
The total nonlocal contribution is then found by summing over all $j \in \cal L'$.

The second difficulty, which we address in this paper, is the 
computational cost of the BS law. It is apparent from Eq.~(\ref{eq:BS}) that
the velocity at one vortex point depends on an integral over all 
other $N-1$ points, so the CPU time required to evolve the vortex
configuration in time scales as $N(N-1) \sim N^2$.
Because of this cost, it has been difficult to numerically
calculate vortex tangles with vortex line density
$L$ (vortex length per unit volume) comparable to what is achieved
in experiments. Similarly, it has been
difficult to perform sufficiently long numerical calculations which are 
needed, for example, to saturate the initial vortex configuration to a 
steady state tangle, or to produce long time series suitable for spectral 
analysis, or to study the decay of quantum turbulence.

Schwarz \cite{Schwarz} proposed the 
Local Induction Approximation (LIA) \cite{DaRios,Arms-Hama}
as a practical alternative to the exact (but CPU-intensive) BS law.
Neglecting the second non-local term in 
Eq.~(\ref{eq:BS_desing}), we have simply

\begin{equation}
\frac{d\bs_i}{dt}=\frac{\kappa}{4\pi}\ln \biggl( \frac{c\langle R \rangle}{a_0} \biggr) \bs_i' \times \bs_i'',
\end{equation}

\noindent
where $c$ is a constant of order unity and $\langle R \rangle $ is the mean 
radius of curvature of the vortex filament. 
The LIA  greatly increases the efficiency of the numerical code because
its computational cost scales with $N$, as opposed to $N^2$ for the
BS law. Moreover the LIA offers insight into vortex dynamics. Using LIA,
it is easy to show
that, in the first approximation, the motion of a point
on the vortex filament is along the direction which is
binormal to the filament at that point, 
with speed which is inversely
proportional to the local radius of curvature $R=1/\vert \bs'' \vert$. 
Moreover the LIA describes
well the translational motion of a circular vortex ring and the rotation
of a small amplitude Kelvin wave.

Unfortunately, the LIA does not capture the interaction between two
vortex filaments,
and the interaction which different parts of the same vortex
have on each other. For example, two circular vortex rings, set coaxially 
close to each other, move in a leapfrogging fashion, but the effect is ignored
by the LIA. Other limitations of the LIA have been discussed in the
literature \cite{Ricca,Boffetta2009,Adachi2010}. Clearly the LIA is
not the best tool to study important properties of turbulence which
arise from the vortex interaction, such as how the energy is distributed 
over the length scales (the Kolmogorov spectrum). 

In a recent paper \cite{tree} we presented the spectrum
of vortex density fluctuations in superfluid turbulence
and compared it to experiments.
This calculation was possible because we used a tree algorithm, which
retains the long-range interaction of the BS law, but sensibly
approximate it, so that its
computational cost scaled as $N\log{(N)}$ rather than $N^2$. The
difference between $N^2$ and $N\log{(N)}$ cannot be overstated: 
it is the same improvement of the Fast Fourier Transform. For example,
in the work reported in ref.~\cite{tree},
 the typical number of vortex points which we used was
of the order of $N \sim 10^5$.
Tree algorithms
are popular in astrophysical N-body 
simulations \cite{Springel2010}. The numerical solution of the gravitational
interaction of $N$ bodies, in fact, has the same difficulty of the 
BS law: the motion
of one body depends on the other $N-1$ bodies, hence the computational cost
scales as $N^2$, as for vortex points.

The aim of this paper is to present a detailed description of a
tree algorithm
tailored to vortex dynamics, provide quantitative error estimates 
associated with the method, and compare results with methods used in the
literature. In particular we
show that the tree method
provides huge computational savings but without the inherent errors
associated with the LIA.

The structure of the article is as follows.
Section \ref{sec:tree} outlines the algorithms and methods required to create a tree structure for a set of 
points embedded in three-dimensional space.
We  pay careful attention to how this tree structure can be efficiently used to compute 
the velocity at a particular vortex point.
In Section \ref{sec:algo} we provide further details on the algorithms required to apply the vortex 
filament method to the study of quantised turbulence.
Section \ref{sec:tree_test} contains the results of numerical tests 
of the tree algorithm.
We show that the $N^2$ scaling inherent in the evaluation of the BS integral 
can be approximated, 
with a tolerable loss of accuracy, by a tree method which exhibits an 
improved $N \log N$ scaling.
A recent, and very detailed, numerical study of counterflow turbulence 
by Adachi \etal \cite{Adachi2010} 
as again demonstrated that the LIA is unsuitable for the study of 
quantum turbulence. 
In Section \ref{sec:counterflow} 
we verify that the tree method does not exhibit the same deficiencies 
of the LIA.

In all simulations presented,
here we use parameters which refer to superfluid $^4$He: circulation
$\kappa=9.97 \times 10^{-4}~\rm cm^2/s$ and vortex core radius
$a_0\approx 10^{-8}~\rm cm$. All our results and methods, however,
can be generalised to vortex systems in low temperature $^3$He-B.


\section{Tree algorithm}\label{sec:tree}
\begin{figure}
  \begin{center}
    \includegraphics[width=0.4\textwidth]{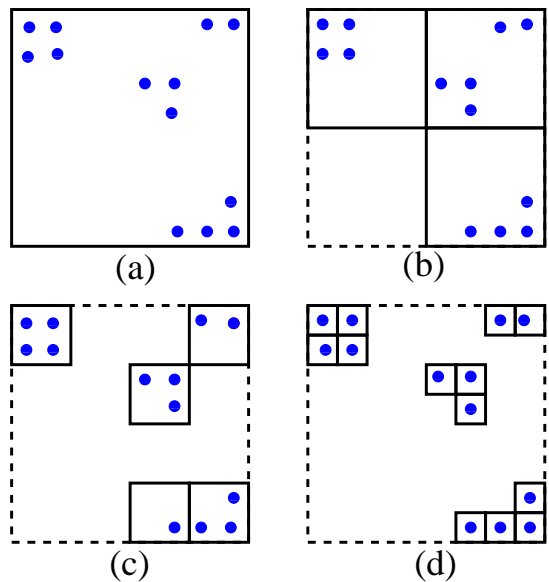}
    \caption{\label{fig:treecartoon}(Color online) Illustration of the tree construction
in two dimensions (quad-tree).
The points (blue dots) are enclosed in the root cell (a),
which is divided into four cells of half size (b), until (c,d) there
is only one point per cell.}
  \end{center}
\end{figure}
Building on the pioneering work of Barnes and Hut \cite{Barnes1986}, 
the majority of astrophysical and cosmological N-body simulations have 
made use of  tree algorithms to enhance the efficiency of the simulation 
with a relatively small loss in accuracy \cite{Bertschinger1998}.
The advantages of these methods are two-fold. Firstly they have been shown 
to scale as $N\log(N))$, where $N$ is the number 
of particles used in the simulation.
Secondly there are now openly available parallel 
implementations of the tree algorithm, which have been used to compute 
simulations with up to 10 billion particles \cite{Springel2005}.
Our implementation of the tree algorithm for vortex dynamics
is limited to a serial program, however the development of a parallel 
scheme is one of our future aims.

We are not the first to apply tree methods to quantised vortex simulations.
Kivotedes \cite{Kivotides2007} used a tree method to enhance a vortex filament simulation, but did not provide any description, testing
and evaluation of the performance of the method.
Kozik and Svistunov \cite{Kozik2005} introduced an interesting method to speed up the simulation
of the Kelvin wave cascade along quantised vortices using a scale separation scheme.
Whilst this is a significant improvement over the full BS integral the problem discussed in \cite{Kozik2005}
is inherently one-dimensional and it is not clear if there is an efficient implementation of the scheme that would
enhance the speed of a three dimensional simulation.

Detailed descriptions of the tree algorithm can be found in a number of 
works \cite{Salmon1991,Dubinski1996}, but their focus is on astrophysical 
simulations. In this paper we
provide a detailed description of the method which is accessible to
the fluid dynamics and the quantum turbulence community.

\begin{figure*}
  \begin{center}
    \includegraphics[width=0.32\textwidth,height=0.3\textwidth]{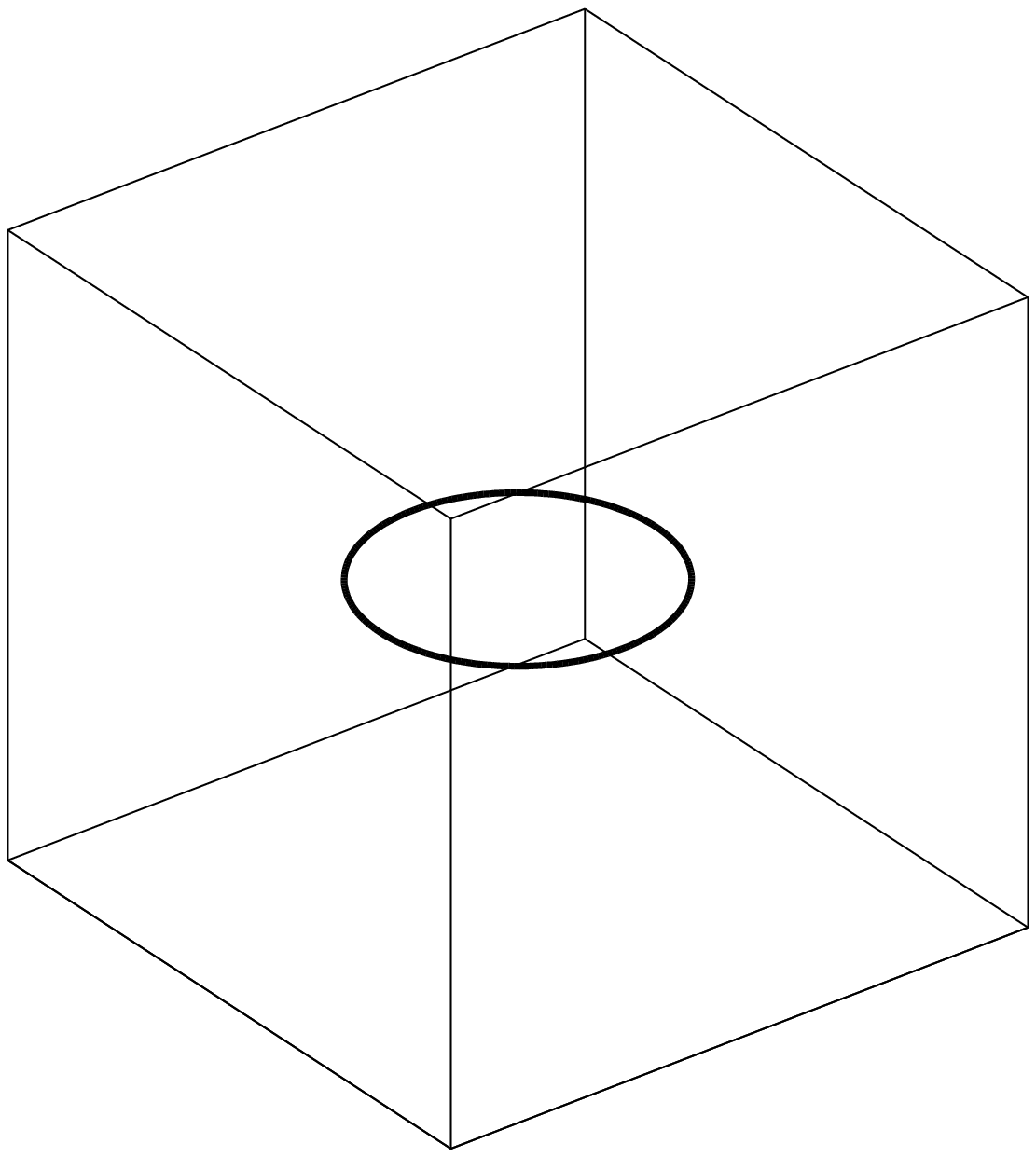}
    \includegraphics[width=0.32\textwidth,height=0.3\textwidth]{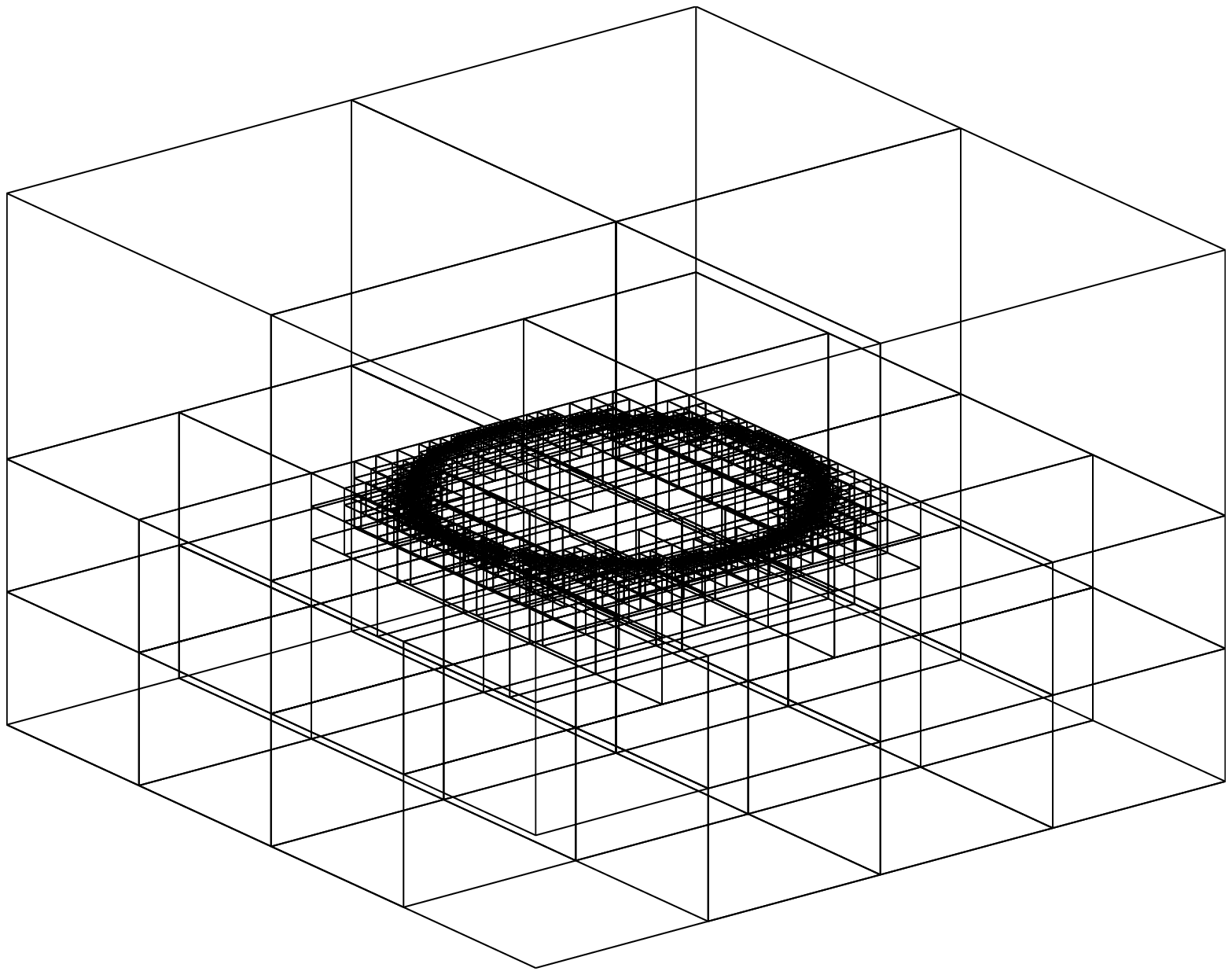}
    \includegraphics[width=0.32\textwidth,height=0.3\textwidth]{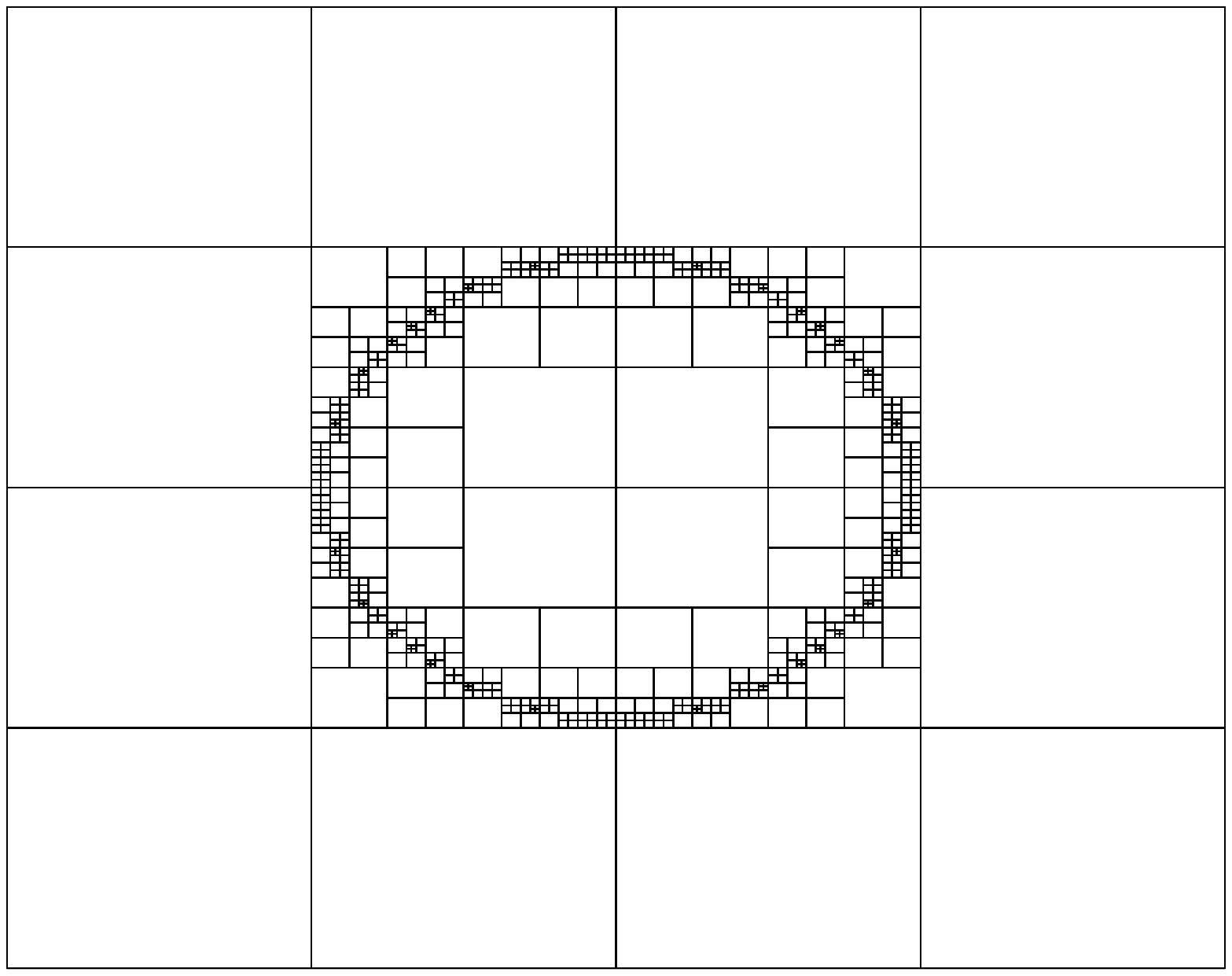}
    \caption{\label{fig:octree}The complex octree structure required to map the vortex points for a single vortex loop (left panel) is plotted in the middle panel. The right panel shows the same structure in the $xy$-plane,at $z=0$.}
  \end{center}
\end{figure*}

At each time-step we have a set of orientated vortex points which 
represent the current position and direction of the vortex filaments.
We first group the vortex points into a hierarchy of cubes which is arranged 
in a three-dimensional octree structure.
The octree can be either be constructed top down or bottom up. We think
that it is conceptually easier to work from the top down,
repeatedly dividing the domain. 
To achieve this aim, we first divide the full computational box, 
the root cell, into eight cubes, and then continue to divide each cube 
into eight smaller `children' cubes,
until either a cube is empty or contains only one vortex point.

The numerical code, {\sc Qvort}\footnote{http://www.staff.ncl.ac.uk/a.w.baggaley/doxy/html/index.html} we have developed is written in FORTRAN 2003 and employs
linked lists \cite{Brainerd} to efficiently create the tree structure.
In the code each cell in the tree is a node which, using pointers, links to the 8 children
within the domain of the parent node.
The main node is set to the full computational box, we then repeatedly call a recursive subroutine to 
allocate the points in this box to the  main nodes child cells.
This process is then repeated dividing the main nodes child cells until the hierarchical tree structure is created.
\begin{figure}
  \begin{center}
    \includegraphics[width=0.5\textwidth,height=0.5\textwidth]{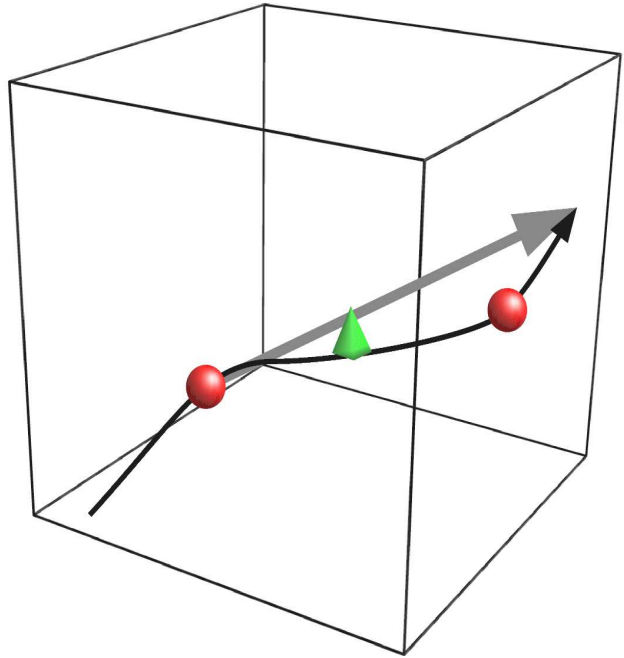}
    \caption{\label{fig:circ_cent} A cartoon of the calculation of the centre of circulation (green) triangle and the total circulation
    bold gray arrow. The (red) circles mark the position of the vortex points within a tree cell and the black line gives a representation of the 
    vortex filament.}
  \end{center}
\end{figure}

As we create the tree, we calculate the total circulation contained within 
each cube and the `center of circulation' using the points that the cube contains.
We denote the number of points the cube contains as $N_{\rm c}$, then the total circulation is given by
\begin{equation}
\bs_\Sigma=\sum_{j=1}^{N_{\rm c}} (\bs_{j+1}-\bs_{j}).
\end{equation}
Likewise the center of circulation is calculated as,
\begin{equation}
\bar{\bs}=\frac{1}{N_{\rm c}}\sum_{j=1}^{N_{\rm c}} \bs_j.
\end{equation}
A schematic diagram for these quantities is displayed in Fig \ref{fig:circ_cent}.

The time required for constructing the tree scales as $N\log(N)$,
so it feasible (and essential) to `redraw' the tree at each time-step.
Fig.~\ref{fig:treecartoon} illustrates this procedure in two dimensions
for clarity.  Fig.~\ref{fig:octree} shows 
the complex nature of the octree required 
for a single vortex ring in a three-dimensional calculation.
 
The enhanced speed of the tree method appears when we  
calculate the induced velocity at each vortex point $\bs_i$.
As opposed to the BS method, we only take the full contribution 
of points near to the vortex point in question.
The induced velocity from far vortex points is an average 
contribution, hence the number of evaluations required per point is 
significantly smaller than $N-1$.
Therefore for each point we must `walk' the tree, and decide if a cell 
in the octree is sufficiently far.

This is done using the concept of the opening angle $\theta$.
In their original work \cite{Barnes1986}, Barnes \& Hut 
calculated $\theta$ by evaluating the distance of the center 
of vorticity from $\bs_i$, which we denote as $d$.
Since each cube in the octree has a known width $w$, then the
opening angle is $\theta=w/d$.
We then must define a critical (or maximum) opening angle 
$\theta_\textrm{max}$.
If $\theta<\theta_\textrm{max}$ we accept the cube, and its contribution 
is used in computing the nonlocal contribution to the velocity via Eq.~(\ref{eq:BS_schwarz_nl}).
The crucial difference between the full BS law and the tree approximation is that the vectors $\mathbf{p}$ and $\mathbf{q}$ must be redefined as $\mathbf{p}=\bar{\bs}-\bs_i$ and $\mathbf{q}=\bs_\Sigma$.
If $\theta_{max}=0$ then each cell contains one point at the
most, and it is clear that $\bs_\Sigma=\bs_{j+1}-\bs_{j}$, $\bar{\bs}=\bs_j$ and we recover $\mathbf{p}$ and $\mathbf{q}$ used in the full BS integral as one would expect.

If $\theta>\theta_{max}$, then we open the cube (assuming it contains more that one 
vortex point) and repeat the test on each of the child cubes that 
it contains.
As a link list is used, from a programming point of view, it is relatively easy to start at the main node and follow the separate branches of the tree, at each stage evaluating whether the cells contribution can be used. The tree-walk ends when the contributions of all cubes have been evaluated.
We note that the adjacent points are exempt from this walk, and their
contribution is taken as in Eq.~(\ref{eq:BS_desing}) so as to avoid the 
singularity.

A correction to this method was proposed by 
Barnes\cite{Barnes1994,Dubinski1996}
to avoid errors which would arise if the center of vorticity 
is near the edge of the cube.
This requires the calculation of the distance from the center of 
vorticity and
the geometrical center of the cube, which we denote as $\zeta$.
The test criteria then becomes $\theta=w/(d-\zeta)$.
In Section \ref{sec:tree_test} we evaluate the improvement caused
by this new criterion.

Whatever opening criterion is used, it is crucial to have a 
reasonable choice for $\theta_\textrm{max}$.
If $\theta_\textrm{max}$ is too large,
the discrepancy between the full BS integral and the tree 
approximation will damage to the accuracy of the simulation.
In Section \ref{sec:tree_test} we shall also show
how the errors in the calculation 
of the velocity at the vortex points depends on $\theta_\textrm{max}$. 

In the next section we describe other necessary details of the 
vortex filament method which we used.

\section{Numerical methods}\label{sec:algo}

The numerical simulations presented here use a $3^{\rm rd}$ order 
Adams--Bashforth scheme. Given an
evolution equation of the form $d\bs/dt=\bv$, the recursion formula
is

\begin{equation}
  \bs_{i}^{n+1}=\bs_{i}^{n}+\frac{\Delta t}{12}(23\bv_{i}^{n}-16\bv_{i}^{n-1}+  5\bv_{i}^{n-2})+\mathcal{O}(\Delta t^4).
\end{equation}

\noindent
where $\Delta t$ is the time-step and the superscript $n$ refers to the 
time  $t_n=n\Delta t$ $(n=0,1,2,\cdots)$.  Lower order schemes are used 
for the initial steps of the calculations, when older velocity values 
are not available.

As the vortex points move, the separation between neighbours along the
same filament is not constant.  In general this is not a problem, 
as we can use finite-difference schemes which work with an adaptive 
mesh size.  However the distance between points sets the resolution of 
the calculation, therefore we must set some upper-bound on the distance 
between vortex points before we re-mesh the filaments
by introducing new vortex points. Our criterion is the following:
if the separation between two vortex points, $\bs_{i}$ and $\bs_{i+1}$,
becomes greater than some threshold quantity $\delta$ (which we call the
minimum resolution), we introduce a new vortex point at 
position $\bs_{i'}$ given by

\begin{equation}
\bs_{i'}=\frac{1}{2}(\bs_i+\bs_{i+1})+\left( \sqrt{R^2_{i'}
-\frac{1}{4}\ell_{i+1}^2}-R_{i'} \right)\frac{\bs_{i'}^{''}}{|\bs_{i'}^{''}|},
\end{equation}

\noindent
where $R_{i'}=|\bs_{i'}^{''}|^{-1}$. In this way
$\bs^{''}_{i'}=(\bs^{''}_i+\bs^{''}_{i+1})/2$,
that is to say the insertion of new vortex points preserves the curvature.
This property is desirable, as the curvature of the filament directly 
affects the velocity field via Eq.~(\ref{eq:BS_desing}). Our procedure
is thus superior to simpler linear interpolation.

Conversely, vortex points are removed if their separation becomes
less than $\delta/2$, ensuring
that the shortest length-scale of the calculation remains fixed. 
This property is important, as the maximum time-step which can be used 
is related to this length-scale.
In the long-wavelength approximation ($ka_0 \ll 1$), the angular frequency
of a Kelvin wave, along a straight vortex is
\begin{equation}
\omega \approx -\frac{\kappa k^2}{4 \pi}\left [ \ln 2/(ka_0) - \gamma\right],
\end{equation}
where $\gamma = 0.5772$ is Euler’s constant.
Hence the timescale for the fastest motions, with a maximum wavenumber $k_\mathrm{max}=4\pi/\delta$, is
\begin{equation}
(\omega_\mathrm{max})^{-1}\approx \frac{\pi(\delta/2)^2}{4\kappa \ln (\delta/2\pi a_0-\gamma)},
\end{equation}
thus a reasonable maximum time-step that can be used is given by
\begin{equation}
\Delta t<\frac{(\delta/2)^2}{\kappa \log (\delta/2\pi a_0)}.
\end{equation}
We note that at finite temperatures this condition may be relaxed as mutual friction will tend to dampen Kelvin waves.
Setting a consistent minimum separation between vortex points makes
the numerical resolution of the calculation transparent, and prevents
the creation of too short length-scales which would considerably
slow down the calculation.

In order to model any turbulent system we must pay particular attention 
to the dissipation of kinetic energy.
Even at very low temperatures, in the absence of mutual friction with
the normal fluid, there are still mechanisms which dissipate the
kinetic energy of quantised vortices:
reconnections, decay of very small
vortex loops and direct phonon emission.  

Using the GPE condensate model, Leadbeater \& al 
\cite{Leadbeater2001} showed
that, at vortex reconnection events, a small amount of kinetic energy
is transformed into acoustic energy of a rarefaction pulse which
moves away. Since our model neglects sound waves and it would be
impractical to determine with accuracy very small changes of kinetic
energy, we interpret this dissipation of kinetic energy as a reduction
of vortex length, essentially using length as a proxy for energy.

We model reconnection events algorithmically in the following way.
If two discretization points, which are not neighbours, become closer 
to each other than the local discretization distance, a numerical 
algorithm reconnects the two filaments by simply switching flags for the 
vortex points in front and behind the filament, subject to the criterion
that the total length decreases.
Self-reconnections (which can arise if a vortex filament has twisted 
and coiled by a large amount) are treated in the same way.
Since reconnections involve only anti-parallel filaments, prior
to reconnection we form local (unit) tangent vectors $\bs'$,
and, using the inner product, 
we check that the two filaments are not parallel.

Clearly, testing for possible reconnections requires a distance 
evaluation between a vortex point and the vortex points in the vicinity.
The most efficient way to do this is within the tree-walk itself,
creating a linked list of all vortex points within a radius of 
$\delta/2$ from a vortex point in question.
This avoids the need to re-loop over all discretization points,
which would involve approximately $N^2$ operations.

We distinguish two kinds of phonon emissions. The condensate model
shows that very small vortex loops, of radius of the order of $a_0$,
which move at speed close to the speed of sound,
are unstable and decay into sound\cite{Jones-Roberts}. It would not be
practical to mesh vortex filaments down to this atomic scale, so
we remove any small vortex loop with less than five discretization
points.

Phonon emission can also be induced by very short Kelvin waves
which rotate rapidly \cite{Vinen2001,Leadbeater2002}. Again, since we do not
mesh filaments down to this scale, we remove vortex points
if the local wavelength is smaller than a specified value 
\cite{tree}. Since our smallest
length-scale is maintained at $\delta/2$, the maximum curvature is 
of the order of $2/\delta$. We proceed in this way: 
at each time-step we compute the local curvature
$C(\xi)=\vert \bs'' \vert$
at each vortex point.
If, at some location along a vortex filament, the local curvature exceeds
the critical value $1.9/\delta~\mathrm{cm}^{-1}$
($95 \%$ of the maximum value $2/\delta$), this vortex point is
removed, smoothing the vortex filament  locally. This loss of length
as a proxy of energy is our model of phonon emission in the low
temperature limit.

We approximate all spatial derivatives using 
$4^{\rm th}$ order finite difference schemes which account 
for varying mesh sizes along the vortex filaments \cite{Gamet1999}.
A high-order scheme is desirable as the spatial derivatives enter 
into the equation of motion via, Eq.~(\ref{eq:BS_desing}). 
Let $\bs_i$ be the $i^{\rm th}$ point on the vortex filament; the two
points behind have positions
$\bs_{i-2}$ and $\bs_{i-1}$, and the two points in front have positions 
$\bs_{i+1}$ and $\bs_{i+2}$.
We denote $\ell_{i-1}=|\bs_{i-1}-\bs_{i-2}|$, $\ell_i=|\bs_{i}-\bs_{i-1}|$, 
$\ell_{i+1}=|\bs_{i+1}-\bs_{i}|$, and $\ell_{i+2}=|\bs_{i+2}-\bs_{i+1}|$.
We have:
\begin{equation}
  \bs_i'=A_{1i}\bs_{i-2}+B_{1i}\bs_{i-1}+C_{1i}\bs_{i}+D_{1i}\bs_{i+1}+E_{1i}\bs_{i+2},
\end{equation}
where the coefficients $A_{1i}$, $B_{1i}$, $C_{1i}$, $D_{1i}$ and $E_{1i}$ are given by,
\begin{equation}
  A_{1i}= \frac{\ell_{i}\ell_{i+1}^2+\ell_{i}\ell_{i+1}\ell_{i+2}}{\ell_{i-1}(\ell_{i-1}+\ell_{i})(\ell_{i-1}+\ell_{i}+\ell_{i+1})(\ell_{i-1}+\ell_{i}+\ell_{i+1}+\ell_{i+2})}
\end{equation}
\begin{equation}
  B_{1i}=\frac{-\ell_{i-1}\ell_{i+1}^2-\ell_{i}\ell_{i+1}^2-\ell_{i-1}\ell_{i+1}\ell_{i+2}-\ell_{i}\ell_{i+1}\ell_{i+2}}{\ell_{i-1}\ell_{i}(\ell_{i}+\ell_{i+1})(\ell_{i}+\ell_{i+1}+\ell_{i+2})}
\end{equation}
\begin{equation}
  C_{1i}=-(A_{1i}+B_{1i}+D_{1i}+E_{1i})
\end{equation}
\begin{equation}
  D_{1i}=\frac{\ell_{i-1}\ell_{i}\ell_{i+1}+\ell_{i}^2\ell_{i+1}+\ell_{i-1}\ell_{i}\ell_{i+2}+\ell_{i}^2\ell_{i+2}}{\ell_{i+1}\ell_{i+2}(\ell_{i}+\ell_{i+1})(\ell_{i-1}+\ell_{i}+\ell_{i+1})}
\end{equation}
\begin{equation}
  E_{1i}= \frac{-\ell_{i+1}\ell_{i}^2-\ell_{i-1}\ell_{i}\ell_{i+1}}{\ell_{i+2}(\ell_{i+1}+\ell_{i+2})(\ell_{i}+\ell_{i+1}+\ell_{i+2})(\ell_{i-1}+\ell_{i}+\ell_{i+1}+\ell_{i+2})}
\end{equation}
Similarly we have:
\begin{equation}
  \bs_i''=A_{2i}\bs_{i-2}+B_{2i}\bs_{i-1}+C_{2i}\bs_{i}+D_{2i}\bs_{i+1}+E_{2i}\bs_{i+2},
\end{equation}
where the coefficients 
$A_{2i}$, $B_{2i}$, $C_{2i}$, $D_{2i}$ and $E_{2i}$ are given by
\begin{equation}
  A_{2i}= \frac{2[-2\ell_{i}\ell_{i+1}+\ell_{i+1}^2-\ell_{i}\ell_{i+2}+\ell_{i+1}\ell_{i+2}]}{\ell_{i-1}(\ell_{i-1}+\ell_{i})(\ell_{i-1}+\ell_{i}+\ell_{i+1})(\ell_{i-1}+\ell_{i}+\ell_{i+1}+\ell_{i+2})}
\end{equation}
\begin{equation}
  B_{2i}= \frac{2[2\ell_{i-1}\ell_{i+1}+2\ell_{i}\ell_{i+1}-\ell_{i+1}^2+\ell_{i-1}\ell_{i+2}+\ell_{i}\ell_{i+2}-\ell_{i+1}\ell_{i+2}]}{\ell_{i-1}\ell_{i}(\ell_{i}+\ell_{i+1})(\ell_{i}+\ell_{i+1}+\ell_{i+2})}
\end{equation}
\begin{equation}
  C_{2i}=-(A_{2i}+B_{2i}+D_{2i}+E_{2i})
\end{equation}
\begin{equation}
  D_{2i}= \frac{2[-\ell_{i-1}\ell_{i}-\ell_{i}^2+\ell_{i-1}\ell_{i+1}+2\ell_{i}\ell_{i+1}+\ell_{i-1}\ell_{i+2}+2\ell_{i}\ell_{i+2}]}{\ell_{i+1}\ell_{i+2}(\ell_{i}+\ell_{i+1})(\ell_{i-1}+\ell_{i}+\ell_{i+1})}
\end{equation}
\begin{equation}
  E_{1i}= \frac{2[\ell_{i-1}\ell_{i}+\ell_{i}^2-\ell_{i-1}\ell_{i+1}-2\ell_{i}\ell_{i+1}]}{\ell_{i+2}(\ell_{i+1}+\ell_{i+2})(\ell_{i}+\ell_{i+1}+\ell_{i+2})(\ell_{i-1}+\ell_{i}+\ell_{i+1}+\ell_{i+2})}
\end{equation}

Note that if we set 
$\ell_{i-1}=\ell_i=\ell_{i+1}=\ell_{i+2}=h$, 
then the above expressions reduce to familiar 
finite-difference schemes for a uniform mesh:
\begin{equation}
\bs_i'=\frac{1}{12h}(\bs_{i-2}-8\bs_{i-1}+8\bs_{i+1}-\bs_{i+2})+\mathcal{O}(h^4),
\end{equation}
\begin{equation}
\bs_i''=\frac{1}{12h^2}(-\bs_{i-2}+16\bs_{i-1}-30\bs_{i}+16\bs_{i+1}-\bs_{i+2})+\mathcal{O}(h^4).
\end{equation}

The final issues to address in this section is the boundary conditions used in the simulations and how they are implemented.
All numerical simulations presented here were performed with
periodic boundary conditions. The computational domain was a cube.
If a vortex point leaves the domain, then it is simply re-inserted 
at the opposite side of the cube. The presence of periodic boundary
conditions affects the calculation of the distance between two vortex
points (two vortex points point at the opposite end of the domain
may be actually close to each other).
We must also use periodic `wrapping' to ensure that the velocity field 
is periodic; in theory we should wrap the cube with an infinite number of 
copies in all directions. In practice, we 
only apply a single layer of wrapping, which still requires a very
large amount of numerical evaluations. If we used the BS law, a single
layer of wrapping would require $27N^2$ evaluations per time-step.
Fortunately again the tree method eases the amount of numerical work,
as the computational domain is wrapped with copies of the octree we 
have already computed. Walking each of these trees, as well as the tree
for the main domain, does not require a significant increase of CPU time.

\section{Tests of the tree algorithm}\label{sec:tree_test}

\begin{figure}
  \begin{center}
    \psfrag{N}{$N$}
    \psfrag{t}{time(s)}
    \psfrag{s1}{$\mathcal{O}(N^2)$}
    \psfrag{s2}{$\mathcal{O}(N \log N)$}
    \includegraphics[width=0.6\textwidth]{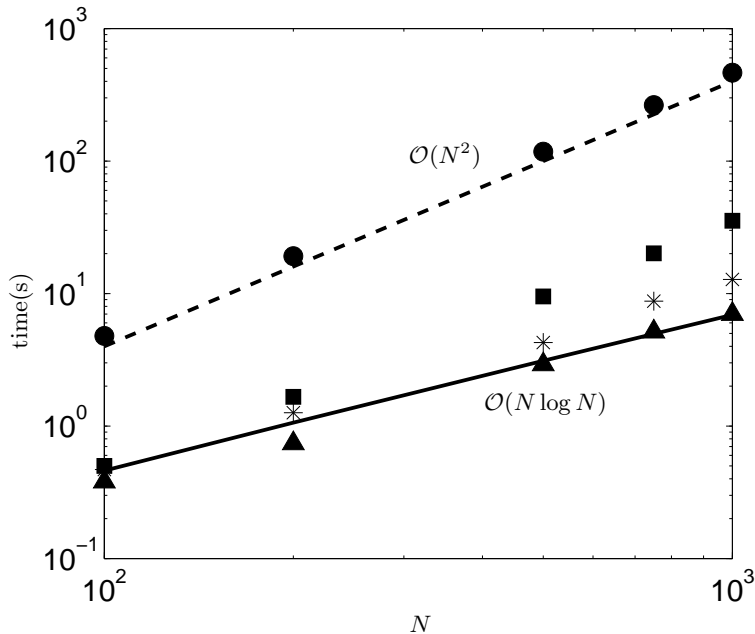}
    \caption{\label{fig:tree_timing} The CPU time required for 
one hundred time-steps, plotted as a function of the number of
vortex points $N$. The calculations were performed on an single processor 
with a clock speed of $2.53~\rm GHz$. The symbols represent different 
values of $\theta_\textrm{max}$ as follows: 
triangles, $\theta_\textrm{max}=0.6$, 
asterisks, $\theta_\textrm{max}=0.4$, 
squares, $\theta_\textrm{max}=0.2$,
and circles, $\theta_\textrm{max}=0.$ (that is, the full BS calculation).}
  \end{center}
\end{figure}

We now test the error associated with the use of the tree approximation,
and show that, with an appropriate choice of $\theta_\textrm{max}$,
the discrepancy between using the tree algorithm and the full BS law is
very small.  All tests are carried out in a periodic cube of side
$0.1~\rm cm$. The minimum resolution is set at $\delta=0.001$cm, and
the time-step is $\Delta t=10^{-5}$s.
For our first test we consider a system of ten random vortex loops with an 
increasing number of vortex points (hence increasing radius).
We determine the CPU time required for evolving the system for
one hundred time-steps as a function of $\theta$; the results are 
presented in Fig.~\ref{fig:tree_timing}. It is apparent that even
for very small maximum opening angles, $\theta_\textrm{max}$, 
there is a huge increase in CPU speed, and for larger values of 
$\theta_\textrm{max}$ the $N\log N$ scaling is apparent.

\begin{figure}
  \begin{center}
    \psfrag{theta}{$\theta_{\textrm{max}}$}
    \psfrag{ebar}{$\bar{\epsilon}$}
    \includegraphics[width=0.6\textwidth]{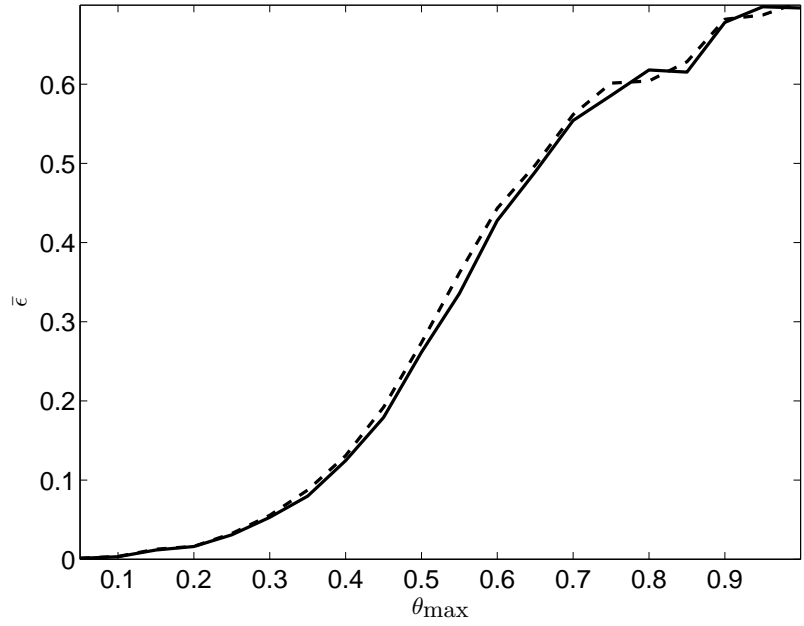}
    \caption{\label{fig:tree_theta} The scaling of the mean relative percentage error in the velocity field, 
$\bar{\epsilon}$, evaluated at discretization points plotted as a function of $\theta_{\textrm{max}}$. 
The dashed line corresponds to the original opening angle criterion
 \cite{Barnes1986} and the solid line 
shows the performance of the correction of Barnes \cite{Barnes1994}.}
  \end{center}
\end{figure}

The second test consists in determining the discrepancies between the 
`true' velocity field (obtained using the BS law) and the velocity
filed obtained with the tree method, as a function of $\theta_\textrm{max}$.
Let $\bv_{i,BS}$ be the
the velocity at the discretization point $\bs_i$ obtained using the 
BS integral, and $\bv_{i,T}$ the corresponding velocity obtained using
the tree method.  We define the mean relative percentage error of the tree 
method as

\begin{equation}
\bar{\epsilon}=100 \times \frac{1}{N}\sum_{i=1}^N \frac{|\bv_{i,BS}-\bv_{i,T}|}{|\bv_{i,BS}|}.
\end{equation}

\noindent
Again, we use a system of 10 randomly oriented and positioned vortex
loops in the same periodic box, each vortex loop discretized into
200 vortex points.
Figure \ref{fig:tree_theta} shows how 
$\bar{\epsilon}$ varies as a function of $\theta_{\textrm{max}}$, for 
both the original tree opening criterion and the correction of Barnes.
Whilst the improvement is modest, since the extra computational cost
is negligible, hereafter we include Barnes's modification.

\begin{figure}
  \begin{center}
    \psfrag{N}{$N$}
    \psfrag{ebar}{$\bar{\epsilon}$}
    \includegraphics[width=0.6\textwidth]{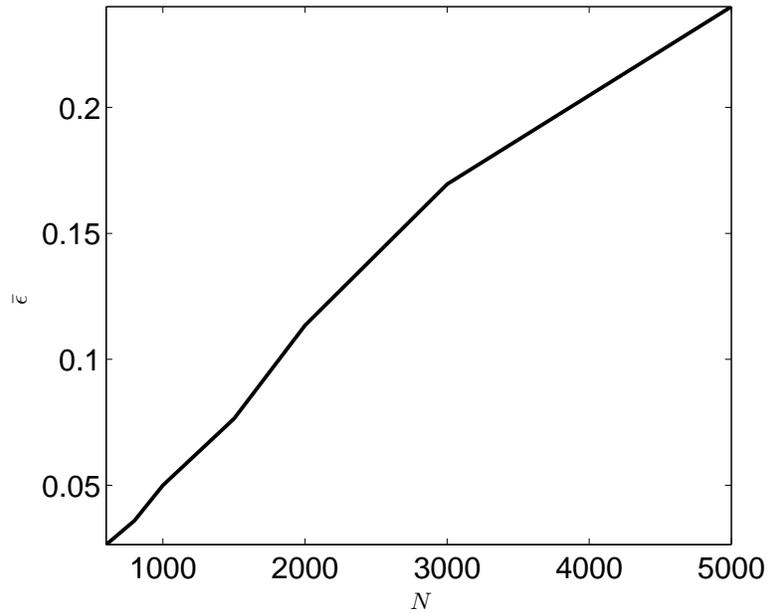}
    \caption{\label{ebarvsN} The mean relative percentage error of the tree approximation for a fixed maximum opening angle, $\theta_{\textrm{max}}=0.4$, with an increasing number of discretization points and therefore vortex length in the system.}
  \end{center}
\end{figure}

The next test investigates how this error varies as a function of $N$ 
for a fixed value of $\theta_{\textrm{max}}$. This test is particularly
important in view of computing very dense vortex tangles, which are not
accessible using the exact BS law.
As before, we consider a system of 10 random vortex loops with an 
increasing number of vortex points in the same periodic box, 
and compute $\bar{\epsilon}$.
Figure \ref{ebarvsN} shows the results for fixed $\theta_\textrm{max}=0.4$.
It is apparent that the error grows with the number of vortex points,
but its magnitude is still small considering the advantages of the tree 
method.  We obtain the same result with a different set up to increase
$N$: we fix the vortex loop size but increase the number of vortex loops.

In the previous tests we have only considered the errors at a single time-step
of a simulation.  It is important to quantify how the difference between
the exact BS law and the tree method varies in time. For this test we
deliberately choose a vortex configuration which the LIA fails to accurately 
model, namely two `leap-frogging' vortices.

\begin{figure*}
  \begin{center}
    \includegraphics[width=0.32\textwidth,height=0.3\textwidth]{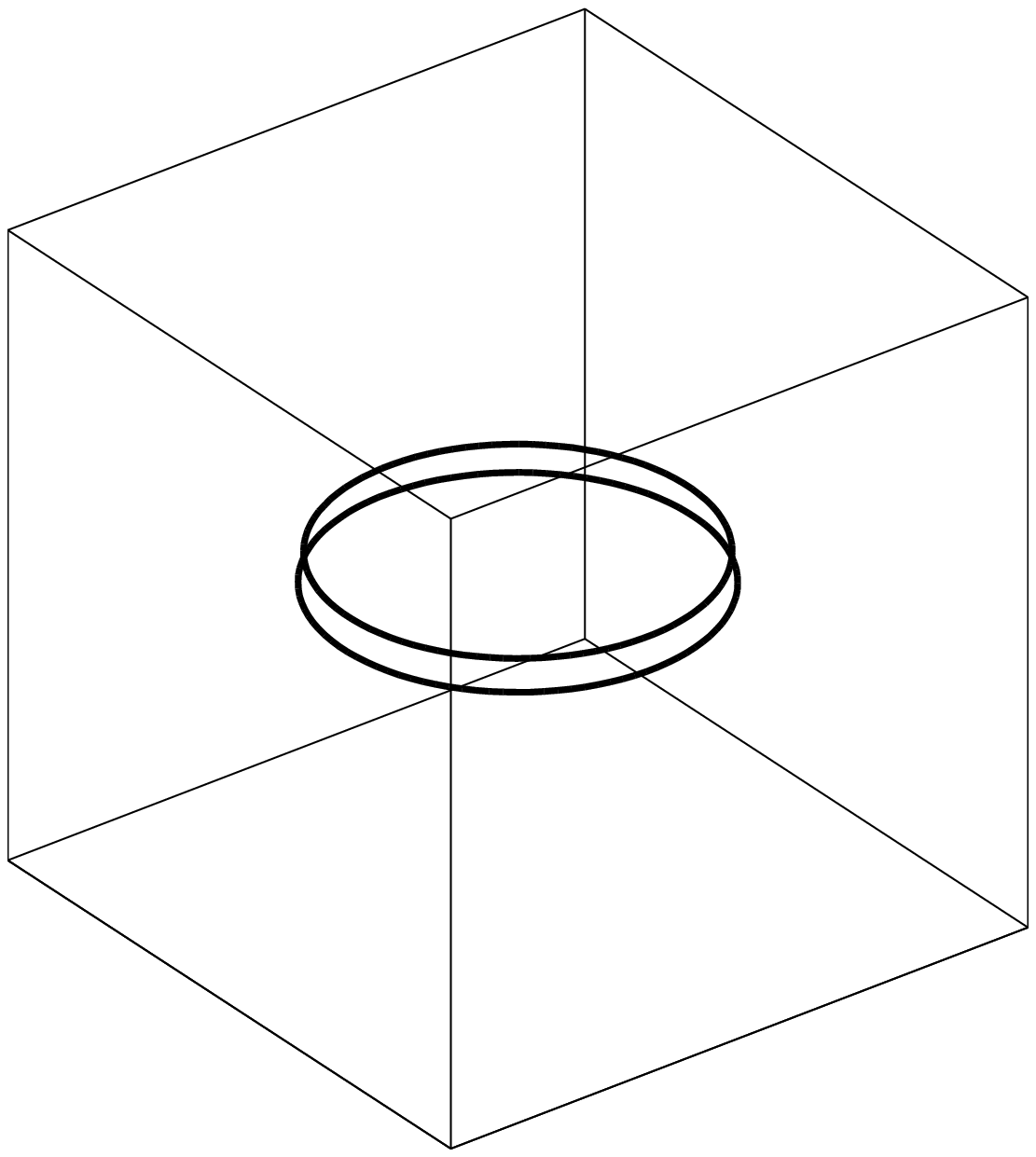}
    \includegraphics[width=0.32\textwidth,height=0.3\textwidth]{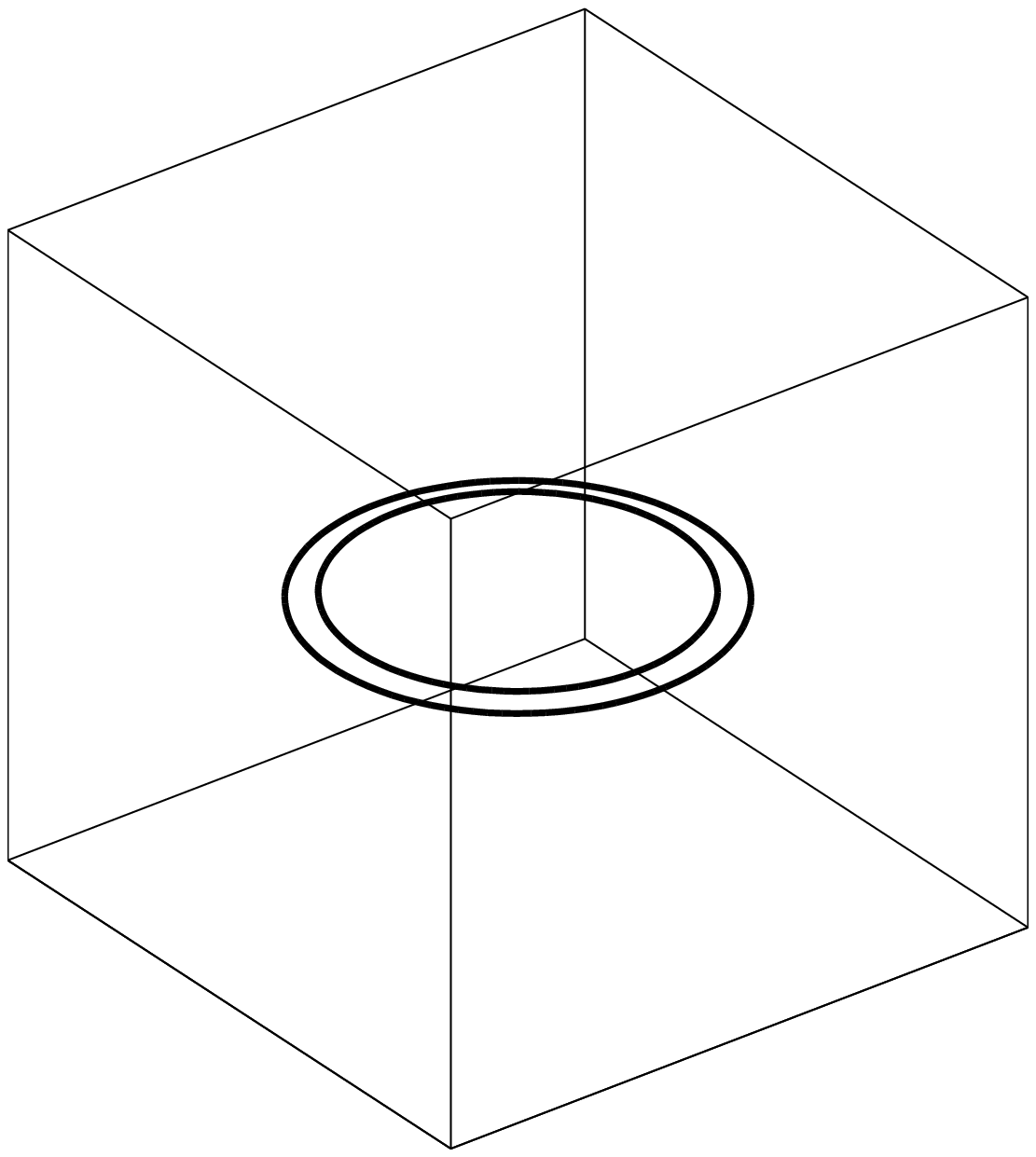}
    \includegraphics[width=0.32\textwidth,height=0.3\textwidth]{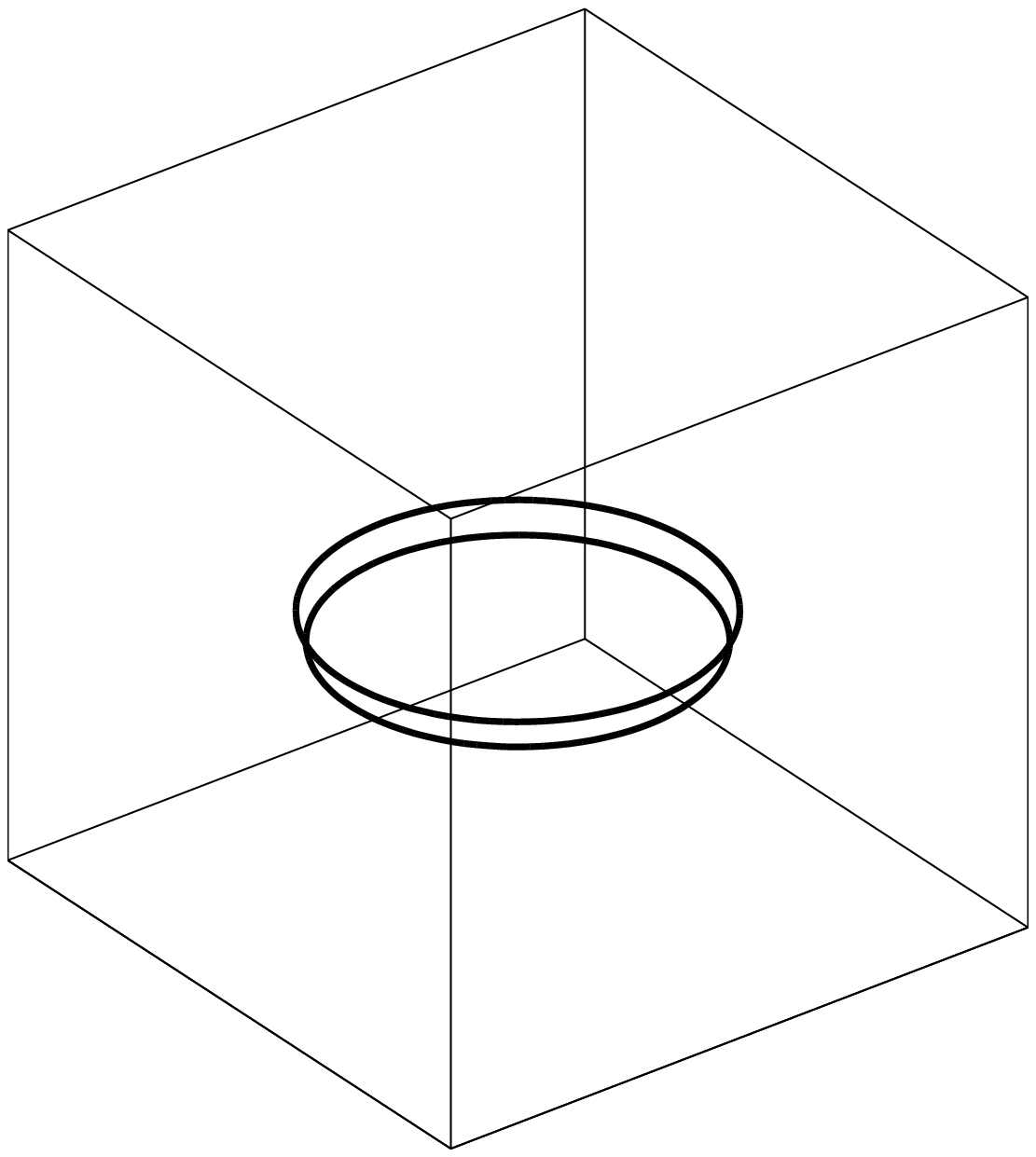}
    \caption{\label{fig:leap_frog} A system of leap-frogging vortices, each loop contains 200 vortex 
points with $\delta=0.001$cm, the mean separation is $0.00075$cm giving each loop a radius of $0.024~\rm cm$. 
These snapshots are taken from a tree code simulation with $\theta_\textrm{max}=0.4$.}
  \end{center}
\end{figure*}

\begin{figure}
  \begin{center}
    \psfrag{t}{$t$}
    \psfrag{diste}{$\bar{D_s}$}
    \includegraphics[width=0.6\textwidth]{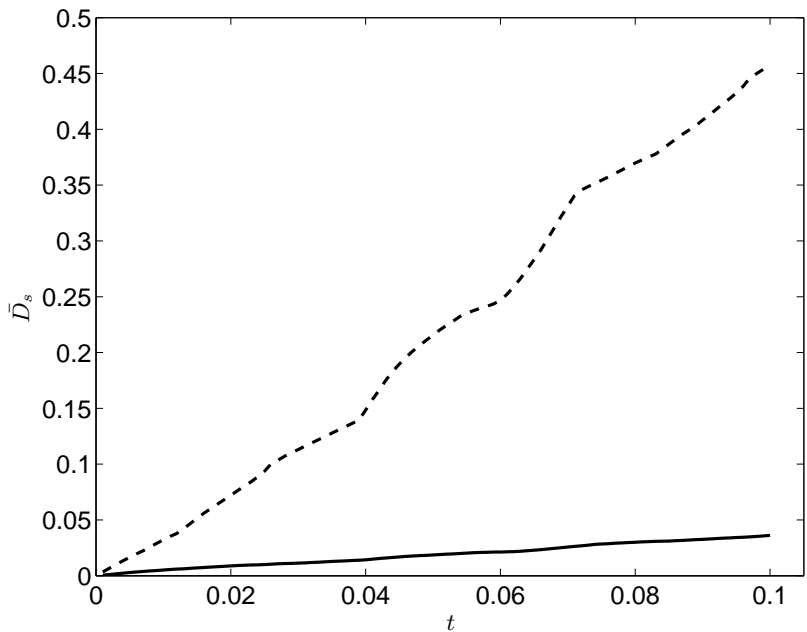}
    \caption{\label{fig:dist_error} The evolution of the normalised distance error statistic $\bar{D_s}$
(see Eq.~(\ref{eq:Dbars})), for a system of leap-frogging vortex rings, plotted as a function of $t$ (seconds). 
The solid line corresponds to a simulation with $\theta_\textrm{max}=0.4$,
and the dashed line to $\theta_\textrm{max}=0.8$.}
  \end{center}
\end{figure}

The initial condition consists of two co-axial vortex rings set
in the $xy$-plane, separated by a small distance in the $z$ direction.
Given the orientation of the vortex lines, the local velocity component -
see Eq.~(\ref{eq:BS_desing}) - drives the vortices  
in the negative $z$ direction.
The non-local velocity component makes the vortex ring which is initially
behind to squeeze through the vortex ring which is in front, overtaking
it; the process repeats over and over again, in a leap-frogging
fashion (note that the vortices maintain the circular shape).
This remarkable behaviour is visible in Fig.~\ref{fig:leap_frog}.
We stress that this behaviour is not captured by the LIA. We repeat
the calculation using the exact BS law, keeping the same number $N$ of vortex
points and the same minimum resolution $\delta$. We then
quantity the difference between the tree method and the BS law by computing

\begin{equation} \label{eq:Dbars}
\bar{D_s}(t)=\frac{1}{N}\sum_{i=1}^N \frac{|\bs_{i,BS}(t)-\bs_{i,T}(t)|}{\delta},
\end{equation}

\noindent
where $\bs_{i,BS}$ is the position of the $i^\textrm{th}$ vortex point in 
the BS calculation, and $\bs_{i,T}$ is the corresponding position in the 
tree method calculation.  The temporal duration of these calculations
captures four separate leap-frogging events.
Figure \ref{fig:dist_error} shows the evolution of $\bar{D_s}$ for 
two different values of $\theta_\textrm{max}$: $0.4$ and $0.8$. 
It is apparent that in both cases the difference between the tree
method and the exact BS evolution is small (smaller
than the minimum discretization level $\delta$), particularly
for $\theta_\textrm{max}=0.4$.

\section{Counterflow simulations}\label{sec:counterflow}

Having benchmarked the tree method for simple vortex configurations,
we change our focus to numerical simulations of quantum turbulence.
We concentrate our attention to turbulence sustained by the relative
motion (counterflow) of normal fluid and superfluid components,
driven by a heat flow at non nonzero temperature. This form of
turbulence which has no analogy in ordinary fluids was
originally studied experimentally by Vinen \cite{Vinen1957}, and,
numerically, by Schwarz \cite{Schwarz}. Counterflow
turbulence
has been recently re-examined by Adachi \& al. \cite{Adachi2010}, who
compared results obtained using the LIA (as in the pioneering
work of Schwarz\cite{Schwarz}) and the full BS law. 
Counterflow turbulence is particularly
suitable for testing the tree method, as Adachi \& al. used it to
discuss the shortcoming of LIA. We want to make sure that the tree
method does not suffer from the same problems of the LIA.

At finite temperatures we must account for the mutual friction between 
the superfluid and the normal fluid.
The equation of motion at position $\bs$ is given by
\begin{equation}
\frac{d\bs}{dt}=\bv_s+\alpha\bs' \times (\bv_n-\bv_s)
-\alpha'\bs' \times \left[ \bs' \times (\bv_n-\bv_s)\right],
\label{eq:Schwarz}
\end{equation}

\noindent
where 
$\alpha$, $\alpha'$ are temperature dependent friction coefficients
\cite{Barenghi1983,Barenghi1998},
$\bv_n$ is the normal fluid's velocity, and the velocity $\bv_s$ is 
calculated using the tree approximation to the de-singularised 
BS integral. As usual, the evolution includes
the algorithmical vortex reconnection procedure.

The calculation is performed in a periodic cube with sides of length 
$0.1~\rm cm$ as done by Adachi \etal \cite{Adachi2010}. 
Superfluid and normal fluid velocities $\bv_n$ and
$\bv_s$ are imposed in the positive and negative $x$ directions
respectively, where $v_{ns}=\vert \bv_n -\bv_s \vert$ is proportional
to the applied heat flux. 
Simulations are performed with the same numerical resolution
$\delta=8 \times 10^{-4}~\rm cm$ and time-step 
$\Delta t=10^{-4}~\rm s$ as used by Adachi \etal
at various value of $v_n$ and $T$. Whereas Adachi \etal used the BS law,
we use our tree method with $\theta_\textrm{max}=0.4$.
The number of vortex points used in these calculations
is not large,  in order to overlap with calculations which used the
BS law:) for example a typical value is $N \approx 6000$ at
$v_{ns}=0.572~\rm cm/s$ and $T=1.9~\rm K$.
It must be stressed that, as in
the works of Schwarz and Adachi \etal, the
back-reaction of the superfluid vortex lines onto the normal fluid
velocity is neglected (an effect which may be significant at low
temperatures). The initial condition consists of six vortex loops
set at random locations, as for Adachi \etal. We find that the results
do not depend on the initial condition.

During the evolution we monitor the vortex line density
\begin{equation}
\label{eq:Ldensity}
L=\frac{1}{V}\int_\mathcal{L} d\xi,
\end{equation}

\noindent
where $V$ is the volume of the computational domain and $\mathcal{L}$ 
is the entire vortex configuration.

\begin{figure*}
  \begin{center}
    \psfrag{t}{$t$}
    \psfrag{L}{$L$}
    \includegraphics[width=0.4\textwidth]{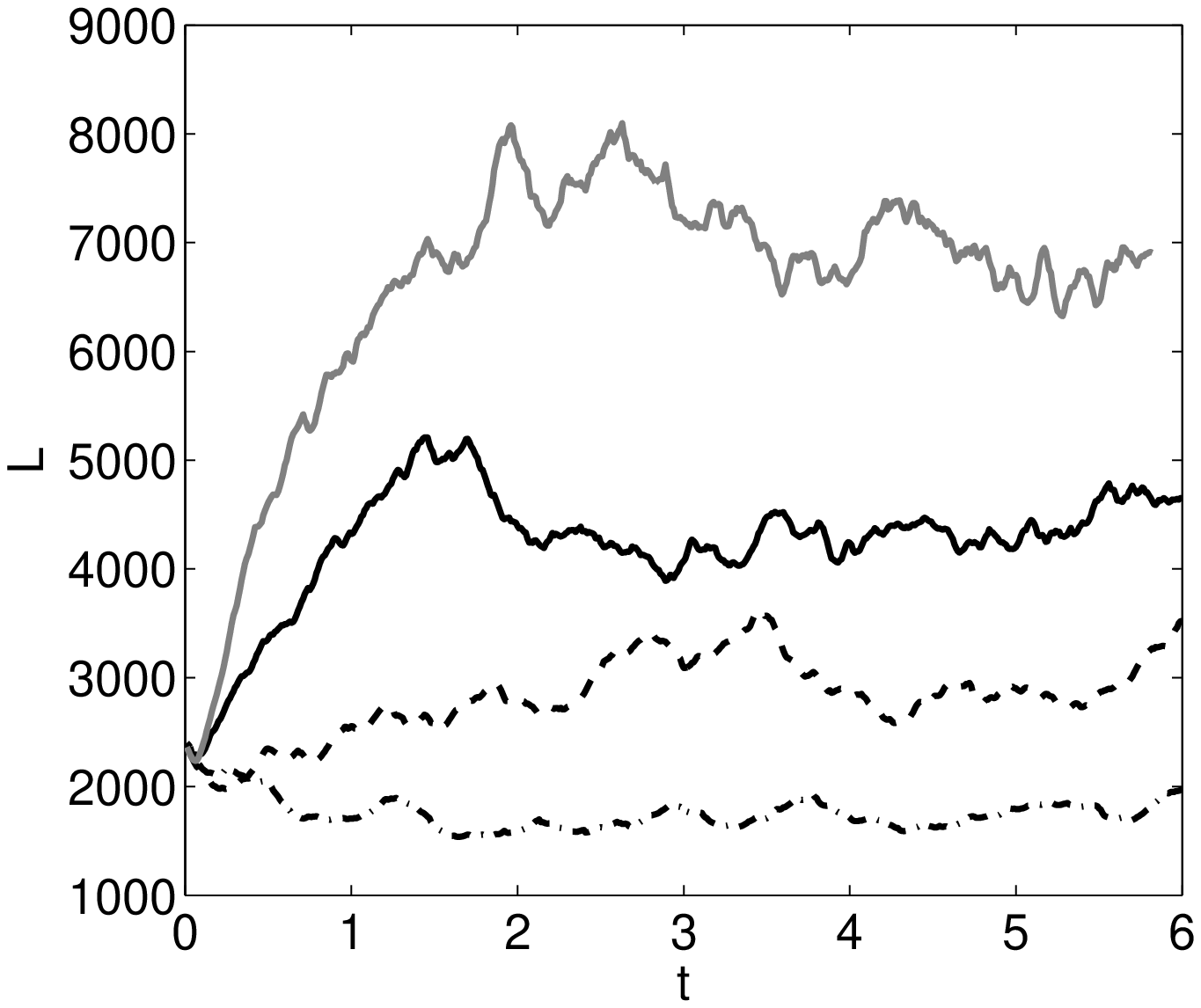}
    \psfrag{L}{$\sqrt{L}$}
    \psfrag{v}{$v_{ns}$}
    \includegraphics[width=0.4\textwidth]{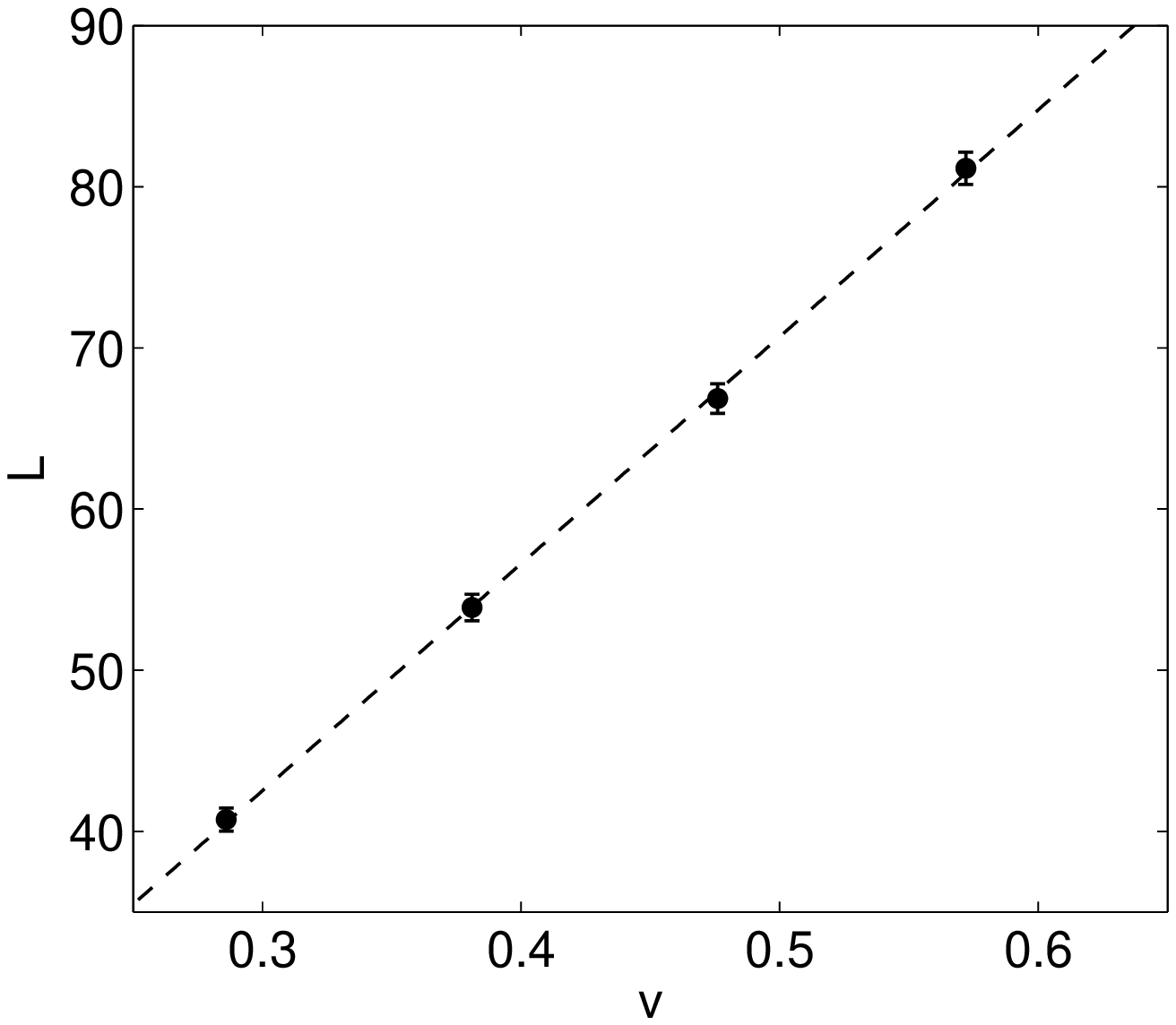}
    \caption{\label{fig:CF_line_density} Left: The evolution of the 
vortex line density $L$ ($\rm cm^{-2}$) vs time $t$ ($\rm s$) for counterflow simulations 
at $1.9~\rm K$ for four counterflow velocities: 
dashed-dotted, $v_{ns}=0.2~\rm cm/s$, 
dashed, $v_{ns}=0.4~\rm cm/s$, 
solid (black), $v_{ns}=0.6~\rm cm/s$,
and  solid (grey), $v_{ns}=0.8~ \rm cm/s$. 
Right: The plot of the vortex line density $L$ ($\rm cm^{-2}$) vs $v_{ns}$ ($\rm cm/s$) confirms
that the saturated line density scales according 
to $L\sim v_{ns}^2$. 
}
  \end{center}
\end{figure*}

We find that, after an initial transient, the vortex line density
saturates to a statistical steady state value, see
Figure \ref{fig:CF_line_density} (left),
which agrees well with the value reported by Adachi \etal
in their Figure 2 at the same temperature and counterflow 
velocity.  
Figure \ref{fig:CF_line_density} (right) shows the expected
relation $\sqrt{L} \sim v_{ns}$ 
between the steady state value of $L$ and the driving counterflow.
More precisely, we obtain $\gamma=140.8$, which compares well with
$\gamma=140.1$ obtained by Adachi \etal, where $L=\gamma^2 v_{ns}^2$.

\begin{figure*}
  \begin{center}
    \psfrag{lpara}{$l_{||}$}
    \psfrag{t}{$t$}
    \includegraphics[width=0.4\textwidth]{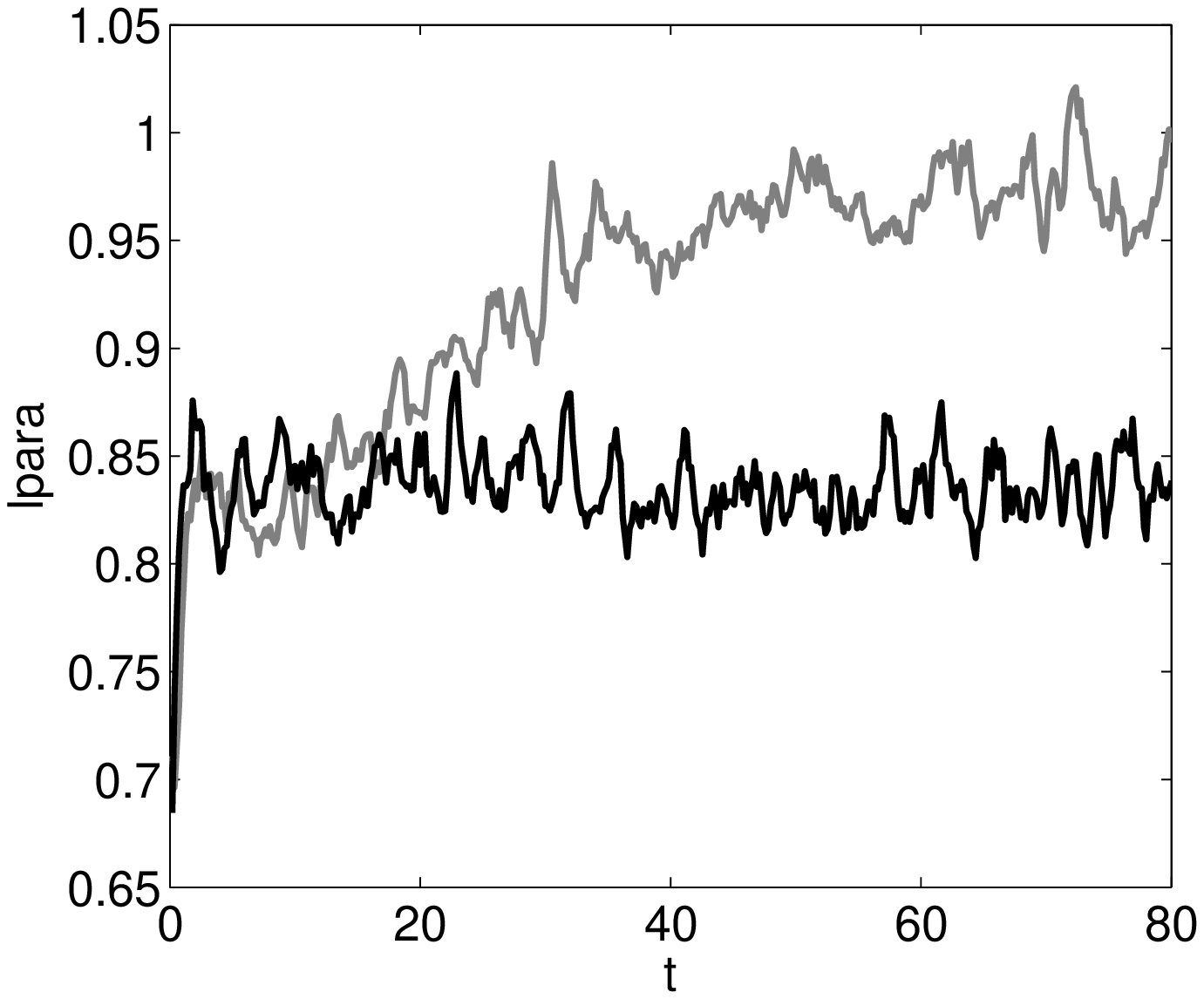}
    \psfrag{lpara}{$l_{||}$}
    \psfrag{v}{$v_{ns}$}
    \includegraphics[width=0.4\textwidth]{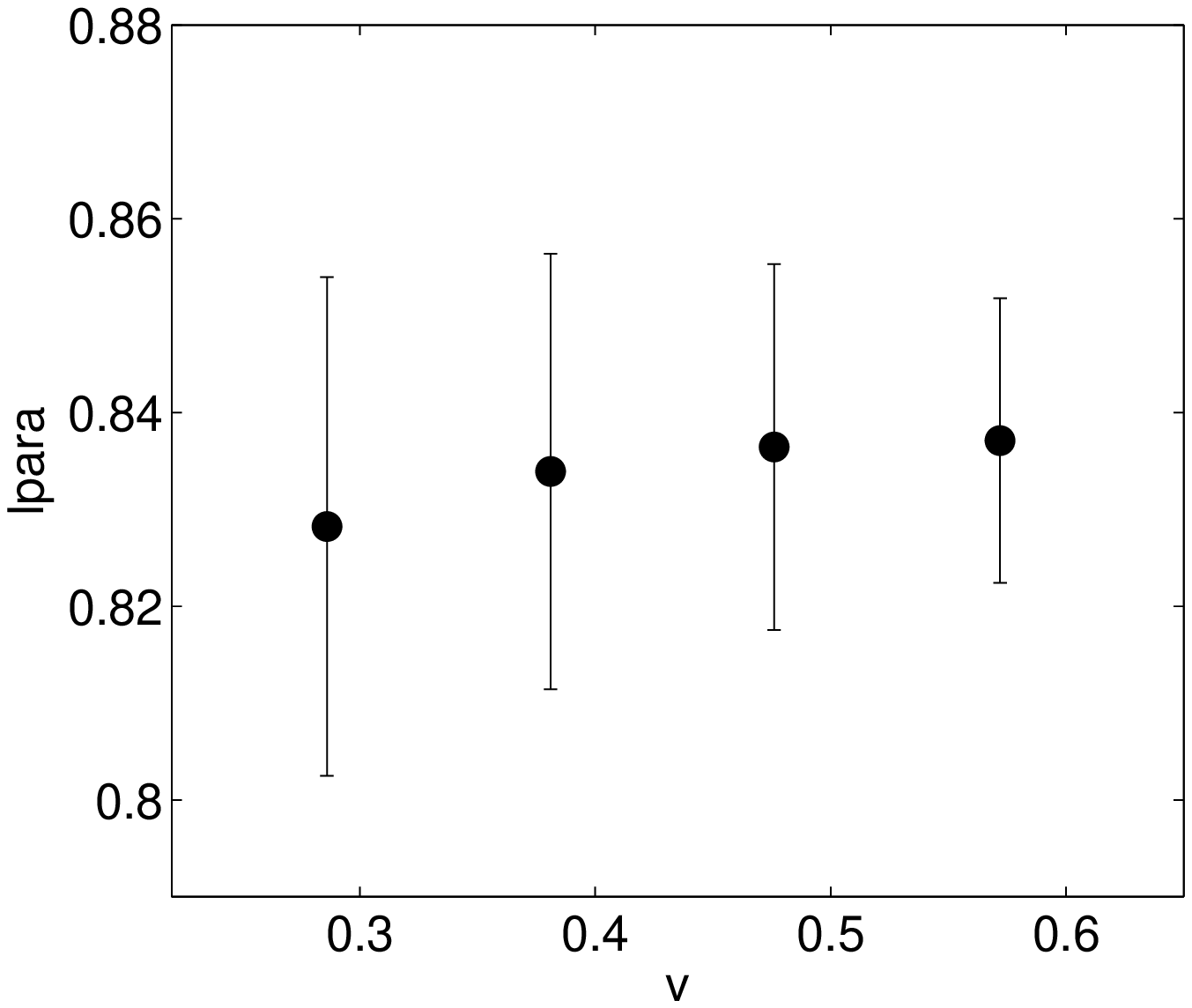}
    \caption{\label{fig:CF_anisotropy}
Left: The anisotropy parameter $l_{||}$ plotted as a function of time $t$ ($s$) for the LIA simulation (grey) 
and for the tree method (black) with $\theta_\textrm{max}=0.4$. 
Right:  The steady-state $l_{||}$ parameter plotted for four different normal fluid velocities 
$v_{ns}$ ($\rm cm/s$) plotted with error bars (calculated from temporal fluctuations) for the tree method only. 
These results are in excellent agreement with those of Adachi \etal\cite{Adachi2010}}
  \end{center}
\end{figure*}

Another important quantity considered by Schwarz and
Adachi \etal
is the anisotropy parameter

\begin{equation}
l_{||}=\frac{1}{VL}\int_\mathcal{L} \left[ 1- (\bs' \cdot \hat{\mathbf{r}}_{||})^2 \right] d\xi,
\end{equation}

\noindent
where $\hat{\mathbf{r}}_{||}$ is the unit vector parallel in the direction 
of the normal fluid.
Figure \ref{fig:CF_anisotropy} (right) shows that $l_{||}$ is almost 
independent of $v_{ns}$, in agreement with experiments and with the
numerical calculations of Adachi \etal
performed using the full BS law (shown in their Fig.~7). 
For example at $v_{ns}=0.55~\rm cm/s$ we have
$I_{||} \approx 0.83$, compared to $0.80$ - $0.85$ of Adachi \etal. 
Figure \ref{fig:CF_anisotropy} also shows that this anisotropy
is not captured by the LIA.

\begin{figure*}
  \begin{center}
    \includegraphics[width=0.35\textwidth]{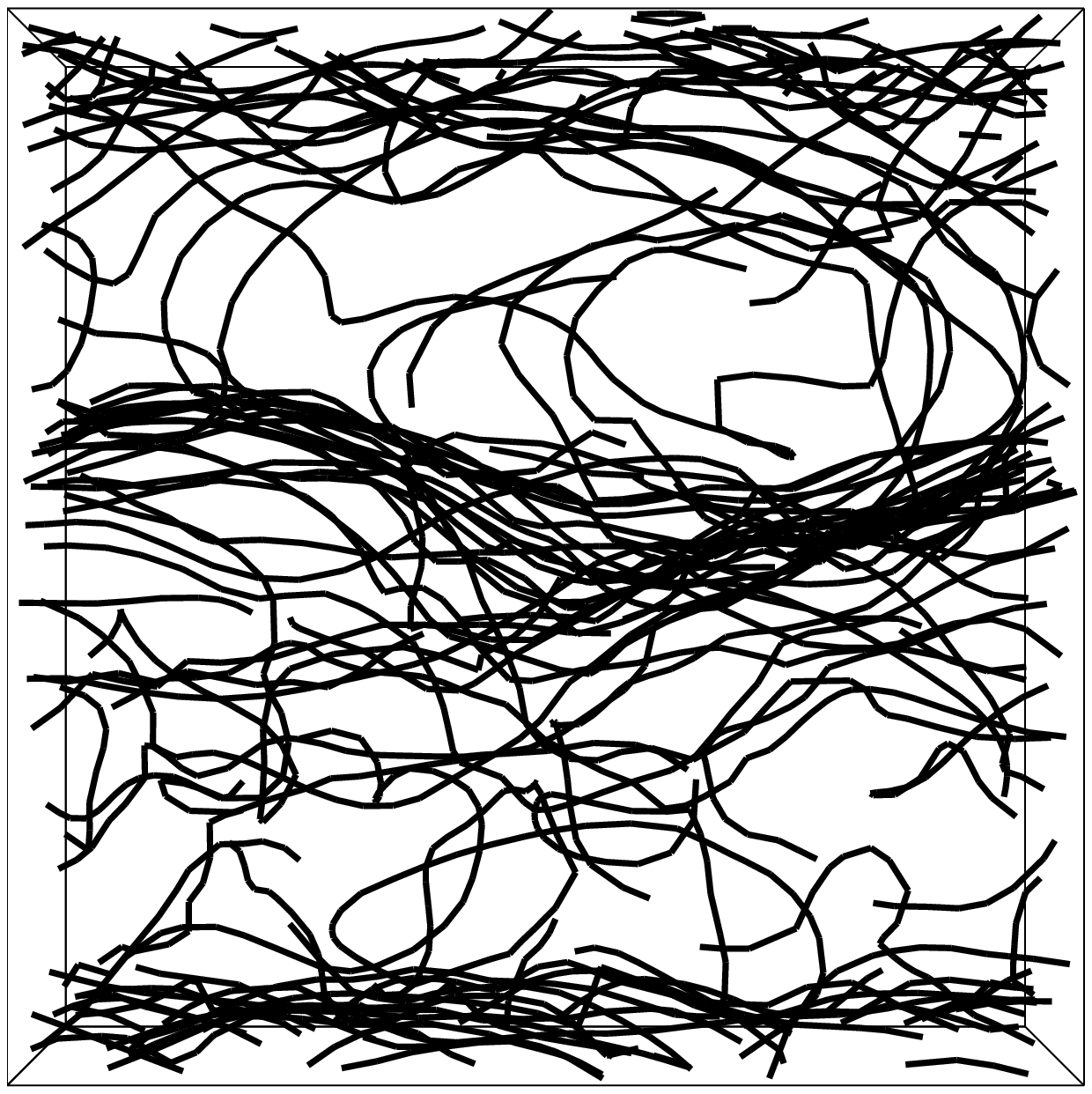}
    \includegraphics[width=0.35\textwidth]{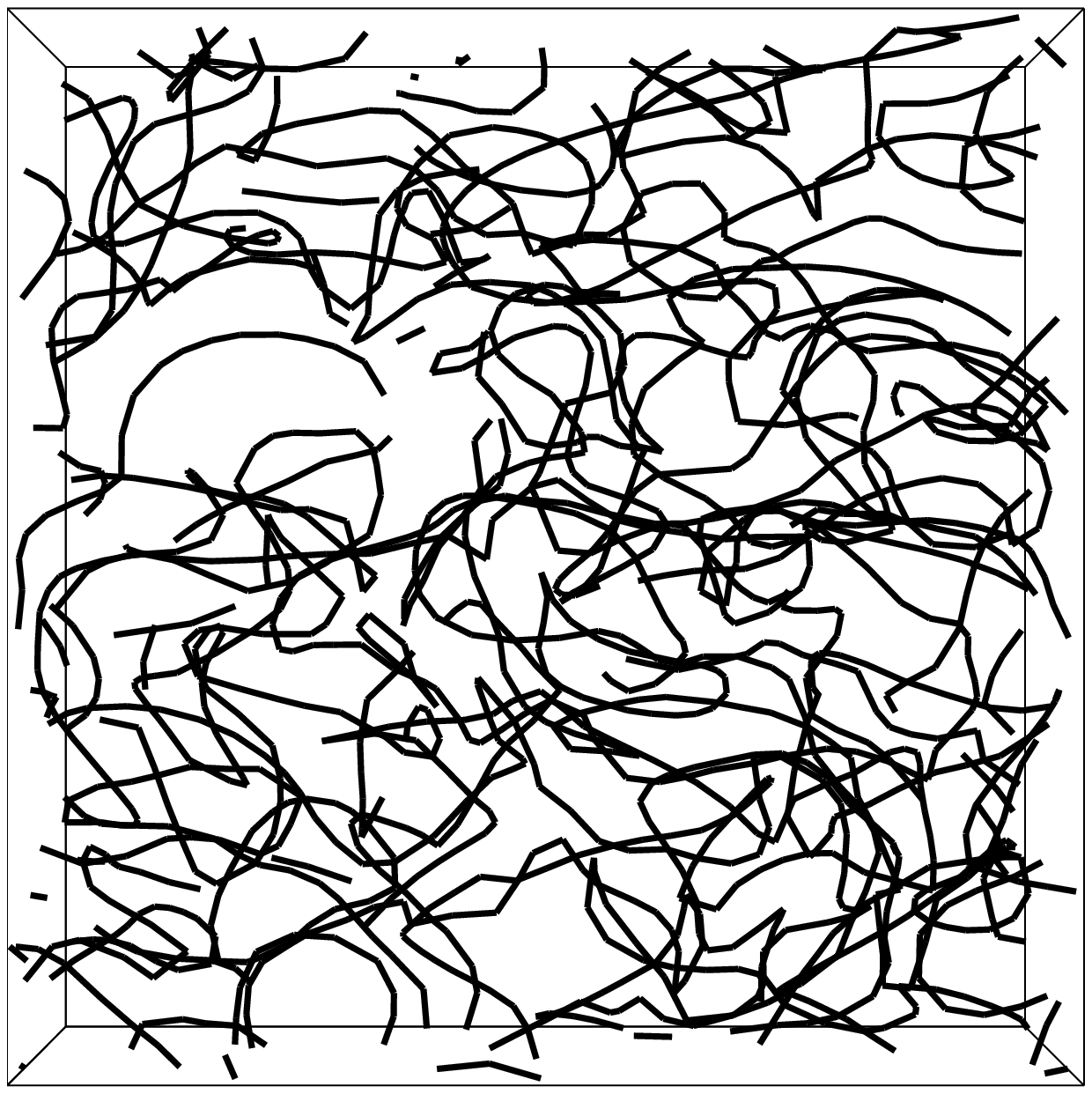}
    \caption{\label{fig:CF_snaps} Views of the tangles from the top of the
computational domain in the $xy$-plane for the LIA simulation (left) and the tree method (right). 
The domain is a cube with sides of length $0.2~\rm cm$; the imposed normal velocity is $v_{ns}=0.5~\rm cm/s$ and $T=1.6~\rm K$.
Note the stratified nature of the tangle wrongly predicted by the LIA.
}
  \end{center}
\end{figure*}

A final qualitative test is shown in Fig.~\ref{fig:CF_snaps}:
the left panel shows the bundling of vortex lines wrongly predicted
by the LIA, and the right panel shows the vortex tangle computed
using the tree method, in agreement with Adachi \etal.


\section{Conclusions}
\label{sec:5}

We conclude that the tree method is an ideal tool to perform numerical
simulations of quantum turbulence. It does not suffer from the known
shortcoming of the LIA, retaining to a good approximation the nonlocal
interaction which is typical of the exact BS law. Unlike the BS law,
whose computational cost scales as $N^2$ where $N$ is the number of
discretization points, the cost of tree method scales as $N\log{(N)}$, thus
making possible numerical calculations of intense vortex tangles for
a sufficiently long time. Finally, the tree
method can be further speeded up by parallelization.


\begin{thebibliography}{}

\bibitem{Donnelly}
R.J. Donnelly, {\it Quantized Vortices In Helium II},
Cambridge University Press, Cambridge (1991)

\bibitem{Barenghi-Sergeev}
C.F. Barenghi \& Y.A. Sergeev (eds.), 
{\it Vortices and Turbulence at Very Low Temperatures},
CISM Courses and Lecture Notes,
Springer (2008)

\bibitem{Halperin}
W. P. Halperin \& M. Tsubota (eds.),
{\it Progress in Low Temperature Physics: Quantum Turbulence}, vol. XVI,
Elsevier (2008)

\bibitem{Tabeling}
J. Maurer \& P. Tabeling,
Europhys. Lett. {\bf 43}, 29 (1998)

\bibitem{Roche2007}
P.-E. Roche, P. Diribarne, T. Didelot, O. Fran\c{c}ais,
L. Rousseau, \& H. Willaime,
Europhys. Lett. {\bf 77},  66002 (2007)

\bibitem{Skrbek}
M. Bla\u{z}kov\'{a}, D. Schmoranzer, L. Skrbek, \& W. F. Vinen
Phys. Rev. B {\bf 79}, 054522 (2009)

\bibitem{Smith1993}
M.R. Smith, R.J. Donnelly, N. Goldenfeld, \& W.F. Vinen,
Phys. Rev. Lett. {\bf 71}, 2583 (1993)


\bibitem{Vinen1957}
W.F. Vinen, Proc. Roy. Soc. A{\bf 240}, 114 (1957);
W.F. Vinen, Proc. Roy. Soc. A{\bf 240}, 128 (1957);
W.F. Vinen, Proc. Roy. Soc. A{\bf 242}, 494 (1957);
W.F. Vinen, Proc. Roy. Soc. A{\bf 243}, 400 (1957)


\bibitem{Paoletti2008}
M.S. Paoletti, M.E. Fisher, K.R. Sreenivasan, and D.P. Lathrop,
Phys. Rev. Lett. {\bf 101} 154501 (2008)

\bibitem{Walmsley2008}
P M. Walmsley \& A.I. Golov,
Phys. Rev. Lett. {\bf 100}, 245301 (2008)

\bibitem{Eltsov2010}
V.B. Eltsov, R. de Graaf, P.J. Heikkinen, J.J. Hosio, 
R. H\"{a}nninen, M. Krusius, \& V. S. L'vov,
Phys. Rev. Lett. {\bf 105}, 125301 (2010)

\bibitem{Lancaster2011}
D.I. Bradley, S.N. Fisher, A.M. Gu\'{e}nault,  R.P. Haley, G.R. Pickett,
D. Potts \&  V. Tsepelin,
Nature Physics {\bf 7}, 473 (2011)

\bibitem{Henn2009}
E.A.L. Henn, J.A. Seman, G. Roati, K.M.F. Magalh\~{a}es, and V.S. Bagnato,
Phys. Rev. Lett. {\bf 103}, 045301 (2009)

\bibitem{Vinen-Niemela}
W.F. Vinen \& J.J. Niemela,
J. Low Temp. Phys. {\bf 128}, 167 (2002) and Erratum,
{\bf 129}, 213 (2002)

\bibitem{Lvov-Nazarenko}
V. S. L'vov, S. V. Nazarenko \& L. Skrbek,
J. Low Temp. Phys. {\bf 145} 125 (2006)


\bibitem{Maryland-tracers}
G.P. Bewley, D.P. Lathrop, \& K.R. Sreenivasan,
Nature {\bf 441}, 588 (2006)


\bibitem{VanSciver}
T. Zhang \& S.W. Van Sciver,
Nature Physics {\bf 1}, 36 (2005)


\bibitem{Lancaster-Andreev}
D.I. Bradley, S.N. Fisher, A.M. Gu\'{e}nault, M.R. Lowe, G.R. Pickett, A. Rahm, \&
R.C.V. Whitehead,
Phys. Rev. Lett. {\bf 93}, 235302 (2004) 


\bibitem{Yale}
W. Guo, S.B. Cahn, J.A. Nikkel, W.F. Vinen, \& D.N. McKinsey
Phys. Rev. Lett. {\bf 105}, 045301 (2010)

\bibitem{Schwarz}
K.W. Schwarz, Phys. Rev. B {\bf 38}, 2398 (1988)

\bibitem{Samuels}
D.C. Samuels,
Phys. Rev. B {\bf 46}, 11714 (1992)
 
\bibitem{Aarts}
A.T.A.M. de Waele and R.G.K.M. Aarts,
Phys. Rev. Lett. {\bf 72}, 482 (1994)

\bibitem{Bauer}
C.F. Barenghi, D.C. Samuels, G.H. Bauer, \& R.J. Donnelly,
Phys. Fluids {\bf 9}, 2631 (1997)

\bibitem{Tsubota2003}
M. Tsubota, T. Araki, \& C.F. Barenghi,
Phys. Rev. Lett. {\bf 90}, 205301 (2003)

\bibitem{Risto}
V.B. Eltsov, A.I. Golov, R. de Graaf, R. H\"{a}nninen, M. Krusius, 
V.S. L'vov, \& R. E. Solntsev,
Phys. Rev. Lett. {\bf 99}, 265301 (2007)

\bibitem{Konda}
L. Kondaurova \& S.K. Nemirovskii,
J. Low Temp. Phys. {\bf 138}, 555 (2005)

\bibitem{Kivo}
D. Kivotides, J.C. Vassilicos, D.C. Samuels, \& C.F. Barenghi,
Phys. Rev. Lett. {\bf 86}, 3080 (2001).

\bibitem{Morris}
K. Morris, J. Koplik, \& D.W.I. Rouson,
Phys. Rev. Lett. {\bf 101}, 015301 (2008)

\bibitem{Kivo2}
D. Kivotides,
Phys. Rev. Lett. {\bf 96}, 175301 (2006)

\bibitem{cascade}
A.W. Baggaley \& C.F. Barenghi,
Phys. Rev. B {\bf 83}, 134509 (2011)

\bibitem{tree}
A.W. Baggaley \& C.F. Barenghi,
Phys. Rev. B {\bf 84}, 020504 (2011)

\bibitem{Kivotides-PIV}
D. Kivotides, C.F. Barenghi, \& Y.A. Sergeev,
Phys. Rev. B {\bf 77} 014527 (2008)

\bibitem{Finne}
A.P. Finne, T. Araki, R. Blaauwgeers, V.B. Eltsov, N.B. Kopnin, 
M. Krusius, L. Skrbek, M. Tsubota, G.E. Volovik, 
Nature {\bf 424}, 1022 (2003)

\bibitem{Saffman}
P.G. Saffman, {\it Vortex Dynamics},
Cambridge University Press (1992)

\bibitem{Koplik}
J. Koplik and H. Levine,
Phys. Rev. Lett. {\bf 71} 1375 (1993)

\bibitem{Bewley}
G.P. Bewley, M.S. Paoletti, K.R. Sreenivasan and D.P. Lathrop,
P.N.A.S. {\bf 105}, 13707 (2008)


\bibitem{Tebbs}
R. Tebbs, A.J. Youd and C.F. Barenghi,
J. Low Temp. Physics {\bf 162}, 314 (2011)


\bibitem{Kerr}
R.M. Kerr
Phys. Rev. Lett. {\bf 106}, 224501 (2011)

\bibitem{Bajer}
M. Kursa, K. Bajer, \& T. Lipniacki,
Phys. Rev. B {\bf 83}, 014515 (2011)

\bibitem{infinite}
C.F. Barenghi, Physica D, {\bf 237} 2195 (2008)


\bibitem{DaRios}
L.S. Da Rios, Rend. Circ. Mat. Palermo {\bf 22}, 117 (1905)

\bibitem{Arms-Hama}
R.J. Arms and F.R. Hama, Phys. Fluids {\bf 8}, 553 (1965)

\bibitem{Ricca}
R.L. Ricca, D.C. Samuels, \& C.F. Barenghi,
J. Fluid Mech. {\bf 391}, 29 (1999)


\bibitem{Boffetta2009}
G. Boffetta, A. Celani, D. Dezzani, J. Laurie \& S. Nazarenko,
J. Low Temp. Phys {\bf 156}, 193 (2009)

\bibitem{Adachi2010}
H. Adachi, S. Fujiyama, \& M. Tsubota,
Phys. Rev. B, {\bf 81}, 104511 (2010)

\bibitem{Springel2010}
V. Springel, 
Ann.  Review  Astronomy and Astrophysics,
{\bf 48}, 391 (2010)

\bibitem{Kivotides2007}
D.Kivotides, Phys. Rev. Lett.,{\bf 76}, 054503 (2007) 

\bibitem{Kozik2005}
E.Kozik \& B. Svistunov, Phys. Rev. B,{\bf 94}, 025301 (2005) 

\bibitem{Barnes1986}
J. Barnes \& P. Hut,
Nature {\bf 324}, 446 (1986)

\bibitem{Bertschinger1998}
E. Bertschinger,
Ann. Review Astronomy and Astrophysics {\bf 36}, 599 (1998)


\bibitem{Springel2005}
V.Springel, S.D.M. White, A. Jenkins, C.S. Frenk,
N.  Yoshida,  L. Gao, J. Navarro,  R. Thacker,
D. Croton, J. Helly, J.A. Peacock, S. Cole,
P. Thomas,  H. Couchman,  A. Evrard,  J. Colberg, \& F. Pearce,
Nature {\bf 435}, 629 (2005)

\bibitem{Salmon1991}
J.K. Salmon,
California Inst.~of Tech., Pasadena (1991)

\bibitem{Brainerd}
W.S.Brainerd, 
{\it Guide to Fortran 2003 Programming}, Springer Verlag, Berlin (2009)

\bibitem{Dubinski1996}
J. Dubinski,
New Astronomy {\bf 1}, 133 (1996)

\bibitem{Barnes1994}
J. Barnes,
{\it Computational Astrophysics}, Springer Verlag, Berlin (1994)

\bibitem{Barenghi1983}
C.F. Barenghi, R.J. Donnelly, \& W.F. Vinen,
J. Low Temp. Phys. {\bf 52}, 189 (1983)

\bibitem{Barenghi1998}
R.J. Donnelly \& C.F. Barenghi,
J. Phys. Chem. Reference Data
{\bf 27}, 1217 (1998)


\bibitem{Leadbeater2001}
M. Leadbeater, T. Winiecki, D.C. Samuels, C.F. Barenghi
\& C.S. Adams, Phys. Rev. Letters {\bf 86}, 1410 (2001)

\bibitem{Jones-Roberts}
C.A. Jones \& P.H. Roberts,
J. Phys. A (Mathematical and General) {\bf 15},  2599 (1982)


\bibitem{Vinen2001}
W.F. Vinen,
Phys. Rev. B {\bf 64}, 134520 (2001)


\bibitem{Leadbeater2002}
M. Leadbeater, D.C. Samuels, C.F. Barenghi and C.S. Adams,
Phys. Rev. A {\bf 67},  015601 (2002)

\bibitem{Gamet1999}
L. Gamet, F. Ducros, F. Nicoud and T. Poinsot,
Int. J.  Numerical Methods in Fluids,
{\bf 29}, 159 (1999)


\end{thebibliography}
\end{document}